\documentclass[twocolumn,amsmath,amssymb,showpacs,letterpaper,superscriptaddress]{revtex4-2}

\usepackage{chemformula}
\usepackage{siunitx}
\usepackage{xspace}
\usepackage{pstricks}
\usepackage{dutchcal}
\usepackage{psfrag}
\usepackage{graphicx}
\usepackage[normalem]{ulem}

\ifdefined\qtyproduct
\else
  \ifdefined\NewCommandCopy
    \NewCommandCopy\qtyproduct\SI
  \else
    \NewDocumentCommand\qtyproduct{O{}mm}{\SI[#1]{#2}{#3}}
  \fi
\fi

\newcommand{\ie}{i.e.\@\xspace}

\renewcommand{\bm}[1]{\boldsymbol{\mathbf{#1}}}
\newcommand{\ud}{\mathrm{d}}
\newcommand{\bra}{\left\langle}
\newcommand{\ket}{\right\rangle}
\newcommand{\mE}{\tilde{E}}
\newcommand{\mI}{\tilde{I}}
\newcommand{\mell}{\tilde{\ell}}
\newcommand{\mg}{\tilde{g}}
\newcommand{\nC}{\mathcal{C}}
\newcommand{\nD}{\mathcal{D}}

\newcommand{\pSHG}{p_{\text{SHG}}}
\newcommand{\gSHG}{g_{\text{SHG}}}
\newcommand{\np}{\mathcal{p}}

\newcounter{tempa}
\newcounter{tempb}
\newcounter{tempc}
\newcounter{tempd}

\newenvironment{ddiag}[1]{\psset{unit=1.3mm,fillstyle=solid,fillcolor=white}
   \begin{pspicture}[shift=-5](0,-7)(#1,7)}{\end{pspicture}
}

\newcommand{\ggmoy}[3]{\psline[linewidth=0.5](#1,#3)(#2,#3)}
\newcommand{\eemoy}[3]{\psline[linewidth=0.5,linestyle=dashed](#1,#3)(#2,#3)}

\newcommand{\gggmoy}[4]{\psline[linewidth=0.5](#1,#3)(#2,#4)}

\newcommand{\pparticule}[2]{\pscircle(#1,#2){1}}
\newcommand{\nnonlineaire}[2]{
   \setcounter{tempa}{#1}
   \addtocounter{tempa}{-1}
   \setcounter{tempb}{#2}
   \addtocounter{tempb}{-1}
   \setcounter{tempc}{#1}
   \addtocounter{tempc}{1}
   \setcounter{tempd}{#2}
   \addtocounter{tempd}{1}
   \psframe(\value{tempa},\value{tempb})(\value{tempc},\value{tempd})
}

\newcommand{\ccorreldeuxc}[4]{
   \psline[linestyle=dashed](#1,#2)(#3,#4)
}

\newenvironment{ddddiag}[1]{\psset{unit=1.3mm,fillstyle=solid,fillcolor=white}
   \begin{pspicture}[shift=-8](0,-10)(#1,10)}{\end{pspicture}
}

\begin{document}

\title{Speckle decorrelation in fundamental and second-harmonic light scattered from nonlinear disorder}

\author{Rabisankar Samanta}
\affiliation{Nano-optics and Mesoscopic Optics Laboratory, Tata Institute of Fundamental Research,
             1 Homi Bhabha Road, Mumbai, 400 005, India}
\author{Romain Pierrat}
\email{romain.pierrat@espci.psl.eu}
\affiliation{Institut Langevin, ESPCI Paris, PSL University, CNRS, 1 rue Jussieu,
             75005 Paris, France}
\author{Rémi Carminati}
\affiliation{Institut Langevin, ESPCI Paris, PSL University, CNRS, 1 rue Jussieu,
             75005 Paris, France}
\affiliation{Institut d’Optique Graduate School, Université Paris-Saclay, F-91127 Palaiseau, France}
\author{Sushil Mujumdar}
\email{mujumdar@tifr.res.in}
\affiliation{Nano-optics and Mesoscopic Optics Laboratory, Tata Institute of Fundamental Research,
             1 Homi Bhabha Road, Mumbai, 400 005, India}

\date{\today}

\begin{abstract}
   Speckle patterns generated in a disordered medium carry a lot of information despite the apparent complete
   randomness in the intensity pattern. When the medium possesses $\chi^{(2)}$ nonlinearity, the speckle is sensitive to
   the phase of the incident fundamental light, as well as the light generated within. Here, we examine the speckle
   decorrelation in the fundamental and second-harmonic transmitted light as a function of varying  power in the
   fundamental beam. At low incident powers, the speckle patterns produced by successive pulses exhibit strong
   correlations, that decrease with increasing power. The average correlation in the second-harmonic speckle decays
   faster than in the fundamental speckle. Next, we construct a theoretical model, backed up by numerical computations,
   to obtain deeper physical insights on the faster decorrelations in the second-harmonic light. Whilst providing
   excellent qualitative agreement with the experiments, the model sheds important light on the contribution of two
   effects in the correlations, namely, the generation of second-harmonic light, and the propagation thereof. 
\end{abstract}

\maketitle

Wave transport in a random medium is a universal phenomenon that transcends the boundaries of various sub-topics such as
optics, condensed matter physics, acoustics, or quantum matter etc \cite{ishimaru1978}. Among all of these, transport of
optical waves has attracted most attention due to the sophisticated experimental capabilities offered by optics. Indeed,
the study of photon transport through disordered media has revealed important facets of transport in all regimes of
disorder, from weak scattering occurring in media like fog to strong scattering in dense powders. With increasing
disorder, incident waves experience multiple scattering where the transport of intensity is described as a diffusion process.
Further increase in disorder leads to exotic phenomena such as weak localization and strong localization are manifest in
the system, which essentially represent reduced or arrested photon transport~\cite{wiersma2013}. Traditionally, all these phenomena were studied in the
linear regime due to the inherent non-interacting nature of photons. However, interactions can be created by introducing
nonlinearities into the media. Materials that respond to higher powers of incident electric fields can be exploited to
create disordered systems that favour nonlinear propagation. The consequences of nonlinearity on the physics of light
transport in disorder have been extensively addressed both in $\chi^{(3)}$ media, that are media exhibiting
intensity-dependent refractive index~\cite{conti2007,shadrivov2010,mafi2017,sharabi2018}, and in $\chi^{(2)}$ media,
that can generate second-harmonic frequencies of light~\cite{agranovich1988, yoo1989, faez2009,valencia2009,savo2020,
muller2021, morandi2022}. In the latter scenario, research efforts have been focused on fundamental physics of diffusion and weak
localization in $\chi^{(2)}$ disorder~\cite{agranovich1988, yoo1989, faez2009,valencia2009}, and on the applicability of
disorder in enhancing nonlinear generation~\cite{savo2020, muller2021, morandi2022}.

\begin{figure*}
   \centering
   \includegraphics[width=0.9\linewidth]{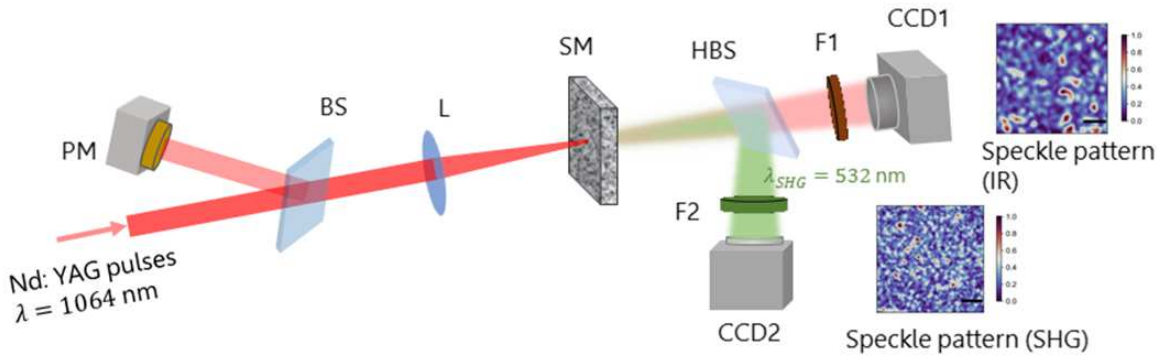}
   \caption{Schematic setup of the experiment. Notations, BS: Beam Splitter, PM: Power Meter, L: Lens , SM: Scattering Medium,
   HBS: Harmonic Beam Splitter, F1: Laser line filter at $\lambda=\SI{1064}{nm}$, F2: Laser line filter at
   $\lambda=\SI{532}{nm}$, CCD1: Charged Coupled Device with InGaAs detector, CCD2: Charged Coupled Device with
   Silicon detector. Two experimental speckle patterns for fundamental (hereafter referred to as IR,
   $\lambda=\SI{1064}{nm}$) and second harmonic generated (hereafter referred to as SHG, $\lambda=\SI{532}{nm}$) light
   are shown here while the adjacent color bars indicate the normalized intensity scale.}
   \label{fig:1}
\end{figure*}

One of the most fundamental effects in disorder that depends on the phase of the propagating light is the appearance of
speckle. A speckle pattern is the random intensity distribution of bright and dark spots developed due to the interference
of many coherent wavelets with the same frequency and different amplitudes and phases travelling in a disordered
medium~\cite{goodman2007}. Despite the apparent complete randomness in the intensity distribution, various
correlations~\cite{feng1988} are known to exist in the speckle pattern. For instance, the optical memory effect
'remembers' the incoming wavefront under slight perturbation in position and
angle~\cite{freund1988,schott2015,judkewitz2015,osnabrugge2019, liu2019}, an idea that has emerged as an efficient tool
in imaging through opaque media~\cite{bertolotti2012,katz2014}. Recent theory and experiments have unveiled non-Gaussian
and long-range correlations between transmitted and reflected speckle patterns~\cite{fayard2015, fayard2018,
starshynov2018}. Not surprisingly, the rich physics of speckle correlations has already motivated research in nonlinear
systems. For instance, a nonlinear optical memory effect~\cite{fleming2019} was revealed in a $\chi^{(3)}$ medium,
namely a silica aerogel, through a series of pump and probe experiments wherein disordered medium is agitated by an
optothermal nonlinearity. Another well-known consequence of $\chi^{(3)}$ nonlinearity is the speckle instability,
wherein the speckle pattern fluctuates and becomes unstable when the nonlinearity surpasses a threshold value
\cite{skipetrov2000,skip2003, skipetrov2004}. In the weak localization regime, the speckle patterns formed by nonlinear
point scatterers exhibit a dynamic instability and lead to chaotic behavior of the system \cite{gremaud2010}.  Such
speckle instabilities in $\chi^{(3)}$ nonlinear disordered media have been experimentally
reported~\cite{smolyaninov2010}. On the other hand, $\chi^{(2)}$ nonlinearity has been employed to primarily investigate
angular correlations in speckles.  For example, experiments and calculations have shown that angular correlations in
reflected speckle scale with sample thickness for second-harmonic light, in contrast to scaling with mean free path for
fundamental light \cite{boer1993}. In another study ~\cite{ito2004}, angular correlations in second-harmonic speckle
under dual-beam excitation were presented in a medium of \ch{LiNbO_3} microcrystals.

In this article, we report our experimental and theoretical studies on intensity-dependent decorrelation in speckle
patterns produced by a second-order nonlinear disordered medium. Specifically, we show that the fundamental and second
harmonic speckle patterns produced by successive incident pulses exhibit strong correlations at low input power which
drop at higher power. The correlation between fundamental speckle patterns remains high compared to the second harmonic.
The decay rate of the average correlation with increasing power is larger in the second harmonic speckles than in the
fundamental speckles. To understand the decorrelation process, we build a theoretical model that traces the propagation
of the linear field, followed by the conversion to second harmonic, finally followed by the propagation of the second
harmonic. The model is in excellent qualitative agreement with the experimental results. The theoretical model is also
backed up with a Monte Carlo computation, which sheds light on two contributions to the decorrelation process, namely,
the decorrelation during the generation of the second-harmonic, and that due to the propagation thereof.

\section{Experiments}

\subsection{Experimental Setup}

Towards the experiment, commercially available KDP (Potassium Dihydrogen Phosphate, $@$ EMSURE ACS) crystal grains were
adopted as our nonlinear material. Initially, the grain sizes ranged from $\sim\text{\SIrange{2}{3}{mm}}$ and were
uneven in shape. The grains were subjected to a ball milling process creating a fine powder of KDP, with particle sizes
ranging from \num{2} to \SI{8}{\micro\meter}. The distribution of grain sizes approximately followed a log-normal
distribution with a peak at \SI{3.11}{\micro\meter} and variance of \SI{1.25}{\micro\meter}. For the speckle
measurement, we prepared two opaque slabs (thickness $\sim\text{\SI{510+-15}{\micro\meter}}$ and
$\sim\text{\SI{680+-20}{\micro\meter}}$) of KDP micro-crystals, and the slabs were sandwiched between two microscopic
slides of thickness $\sim\text{\SI{170+-5}{\micro\meter}}$. A coherent backscattering
(CBS)~\cite{wolf1985,akkermans1986} experiment estimated the transport mean free path ($\ell_t$) of the slabs, and the
estimated values were approximately \SI{352}{\micro\meter} and \SI{169}{\micro\meter} at $\lambda=\SI{1064}{nm}$ and
$\lambda=\SI{532}{nm}$ respectively. Fig.~\ref{fig:1} illustrates the schematic of the experimental setup for the
speckle correlations measurements.

\ch{Nd:YAG} laser pulses (EKSPLA, PL2143B, pulse width $\sim\SI{30}{ps}$) with the fundamental wavelength of
$\lambda=\SI{1064}{nm}$ (hereafter referred to as IR), were chosen as our input beam. A glass wedge was introduced in
the incident path to direct a small fraction ($\sim\SI{4}{\%}$) of the beam to a power meter (PM, Ophir Optronics, resolution: $\SI{10}{\micro W}$)
for the input power measurement. The residual beam was then focused onto the scattering medium (SM) through a lens (L) of focal
length \SI{10}{cm}. To avoid damage to the sample, it was placed slightly away from the focus. The transmitted light
consisted of both the fundamental and second harmonic light (here referred to as SHG, $\lambda=\SI{532}{nm}$). A
harmonic beam splitter (HBS) was employed to separate out the two components. The transmitted (IR) and reflected (SHG)
lights from the HBS were then directed to the CCD1 and CCD2 respectively. The CCD1 was an InGaAs detector (SWIR camera,
Photonic Science, UK) with pixel dimension \qtyproduct{30 x 30}{\micro\meter} while the CCD2 was a silicon detector (iXon
Ultra 897, Andor technology) with pixel dimension \qtyproduct{16 x 16}{\micro\meter}. A laser line filter (F1) at
$\lambda=\SI{1064}{nm}$ was added in front of the CCD1 to block any unwanted SHG photon. Similarly, a laser line filter
(F2) at $\lambda=\SI{532}{nm}$ was placed in front of the CCD2. The laser fired at a repetition rate of \SI{1}{Hz}, and
simultaneous measurements of the pulse power and the corresponding IR and SHG speckle patterns were made.

\subsection{Results}

An intense pulse of laser light impacts the disordered sample and imparts certain radiation pressure, which causes the
particles to be displaced from their original position. Overall, the disorder configuration at the input face is
modified, in proportion to the pump power. See Supplemental Document, Sec.~I for details. Since the disorder
configuration changes with every impacting optical pulse, it is imperative to avoid cumulative reconfigurations
happening through multiple pump pulses. Therefore, we only grab two successive speckle patterns in two consecutive pump
pulses, and then translate the sample so as to illuminate a different location on the sample. The homogeneity of
disorder strength was constant across the total area, as also certified by the systematic variation in the results. The
correlation coefficient between two speckle patterns $A$ and $B$ (both $m \times n$ matrices) was calculated as
\begin{equation}\label{correl_expe}
   \nC_{\text{expe}}=\frac{\sum_{i=1}^{m}\sum_{j=1}^{n}(A_{ij}-\overline{A})(B_{ij}-\overline{B})}
      {\sqrt{(\sum_{i=1}^{m}\sum_{j=1}^{n}(A_{ij}-\overline{A})^2)(\sum_{i=1}^{m}\sum_{j=1}^{n}(B_{ij}-\overline{B})^2)}}.
\end{equation}
where the over-bar represents the mean of the matrix. 

\begin{figure}[ht]
   \centering
   \includegraphics[width=0.8\linewidth]{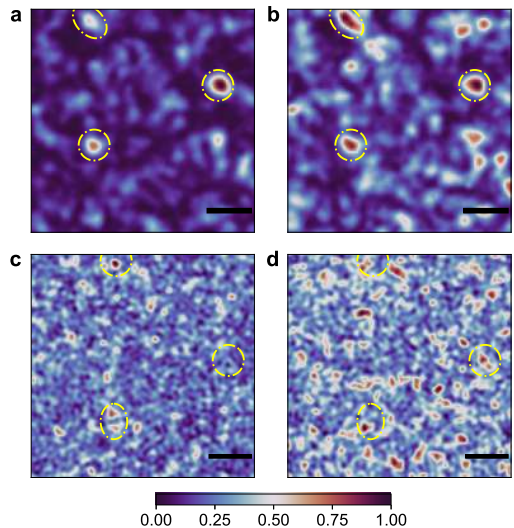}
   \caption{Initial two consecutive speckle patterns of IR [(a) and (b)] and SHG [(c) and (d)] scattered from a
   $\SI{510}{\micro\meter}$ thick sample at an input power of \SI{10.53}{mW}. IR speckle shows higher correlation
   (0.71), and the marked regions with yellow circles emphasize the agreement. On the contrary, the correlation for SHG
   light is observed to be low (0.23), and any regions arbitrarily chosen in the pattern (yellow circles) do not
   show visible agreement. Color bar indicates normalized intensity. The scale bar represents $\SI{600}{\micro\meter}$.}
   \label{fig:2}
\end{figure}

Initial two consecutive speckle patterns of IR and SHG at an input power of \SI{10.5}{mW} are presented in
Fig.~\ref{fig:2}\,(a,b) and (c,d) respectively. Obvious agreement between (a) and (b) is seen, with the yellow circles
emphasizing the regions of clear similarity. For the SHG wavelength, there are no similarities in the speckle patterns
between two consecutive pulses, indicating strong decorrelation within two pulses. The correlation coefficient $\langle
\nC_{\text{expe}} \rangle$ was averaged over 10 sets of speckle patterns, each grabbed at a different location on the
sample at the same pump intensity. 

\begin{figure}[h]
   \centering
   \includegraphics[width=\linewidth]{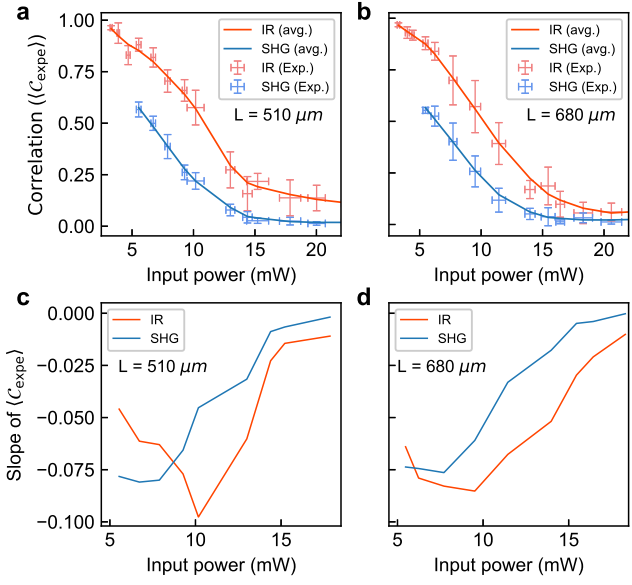}
   \caption{Correlation coefficient between first two consecutive speckle patterns of IR (red markers) and SHG (blue
   markers) light as a function of input power of the fundamental light, calculated from measured speckle patterns. The
   solid lines are obtained after smoothing the experimental data. The data are measured on two different samples with
   thickness $L=\SI{510}{\micro\meter}$ (a) and $L=\SI{680}{\micro\meter}$ (b). Slopes of $\langle \nC_{\text{expe}}
   \rangle_{\text{IR}}$ and $\langle \nC_{\text{expe}} \rangle_{\text{SHG}}$ calculated on each power value for the two
   samples are shown in (c) for $L=\SI{510}{\micro\meter}$ and (d) for $L=\SI{680}{\micro\meter}$.}
   \label{fig:3}
\end{figure}

Figure~\ref{fig:3} reveals the variant decorrelation with pump power for the fundamental and second-harmonic light. A
monotonic decrease in the correlation coefficient is observed in both samples of thicknesses $L=\SI{510}{\micro\meter}$
and $L=\SI{680}{\micro\meter}$. At low powers, up to about 15~mW, the correlation drops rapidly, where-after the rate
reduces with further increase in power. It can be expected to asymptotically approach zero. To compare the qualitative
rate of decorrelation between IR and SHG, we calculate the slopes of the two curves for each point and plot them in
Fig.~\ref{fig:3}\,(c). For $L=\SI{510}{\micro\meter}$ and IR light, the slope initially drops indicating a slowing down
of the decorrelation with increasing power. Subsequently, it rises monotonically. For the SHG light, an initially static
slope is seen to rise monotonically and then saturate at highest power. For the thicker sample, the trends are very
similar. The intersection between the blue and red curves indicates the pump power where the decorrelation rates are
same. Evidently, the two curves intersect at a lower pump power for the thicker sample. The slopes of the two curves
represent the valuable diagnostic for comparing with the theoretical model, which is discussed later.

The source of the fluctuating speckle pattern can be traced to the radiation pressure of the incident pulses, which
induces displacements in the scatterers in random directions. This was experimentally verified in our earlier study
wherein we showed the decrease in speckle contrast with pump power~\cite{samanta2020}. A given absolute displacement of
the scatterers amounts to a smaller relative displacement with respect to the wavelength for the IR light, as compared
to the SHG light. However, the origin of the behavior of decorrelation seen in Figure~\ref{fig:3} is not obvious, and
needs to be rigorously evaluated. This is carried out in the next section.

\section{Theoretical Model}

In parallel to the experiment, we have developed a theoretical model based on coupled transport equations for the linear
($\lambda=\SI{1064}{nm}$) and second harmonic ($\lambda=\SI{532}{nm}$) lights. This model provides
physical insights on the origin of the faster decorrelation for the second harmonic speckle compared to the decorrelation
of the linear speckle. Before deriving the model, we first focus on the correlation function defined in
Eq.~(\ref{correl_expe}). It can also be written as
\begin{equation}\label{correl_expe2}
   \nC_{\text{expe}}=\frac{\int_{\text{CCD}}\delta I(\bm{r})\delta \mI(\bm{r})\ud\bm{r}}
      {\left[\int_{\text{CCD}}\delta I^2(\bm{r})\ud\bm{r}\int_{\text{CCD}}\delta \mI^2(\bm{r})\ud\bm{r}\right]^{1/2}}
\end{equation}
where $\delta I=I-\bar{I}$, $I$ being the intensity and $\bar{I}=\int_{\text{CCD}}I(\bm{r})\ud\bm{r}$. $\mI$ denotes the
intensity once the scatterers have moved due to radiation pressure. It is important to note that this correlation
function does not correspond to the correlation of speckle patterns at different times but measures the correlation
between the speckles produced by two slightly different disorder configurations, the scatterer displacements being
induced by the radiation pressure effect. Assuming ergodicity, we can replace the integration over the pixels of the CCD
camera by a statistical average over all possible disorder configurations which is denoted by $\bra\cdot\ket$. Moreover,
we consider that the statistical properties of the medium are the same after the displacements of the scatterers, \ie
$\bra I\ket=\bra \mI\ket$. Next, we assume that the field has Gaussian statistics (or equivalently that the speckles are
fully developed), which is valid in the regime $k_0\ell_s\gg 1$, $k_0=\omega/c=2\pi/\lambda$ being the wave number and
$\ell_s$ the scattering mean-free path. This implies that $\bra \delta I^2\ket=\bra I\ket^2$.  The correlation function
in Eq.~(\ref{correl_expe2}) becomes $\nC_{\text{expe}}\sim\nC_I-1$ where
\begin{equation}
   \nC_I(\bm{r})=\frac{\bra I(\bm{r})\mI(\bm{r})\ket}
                      {\bra I(\bm{r})\ket^2}.
\end{equation}
Finally, we also have $\nC_I=1+|\nC|^2$ where $\nC$ is the field-field correlation
function given by
\begin{equation}\label{correl_theo}
      \nC(\bm{r})=\frac{\bra E(\bm{r})\mE^*(\bm{r})\ket}
                          {\bra E(\bm{r})E^*(\bm{r})\ket},
\end{equation}
$E$ being the electric field and the superscript $*$ denoting the complex conjugate.  It is important to note that we make the
assumption of a scalar field for the sake of simplicity. This can be justified in the multiple scattering
regime where the field can be considered to be fully depolarized~\cite{VYNCK-2014}. We finally have
\begin{equation}
   \nC_{\text{expe}}\sim|\nC|^2.
\end{equation}
The problem of estimating theoretically $\nC_{\text{expe}}$ now reduces to the computation of $\nC$ for two different frequencies,
\ie $\omega$ for the linear beam and $2\omega$ for
the second harmonic one. The purpose of the next subsections is to develop a transport model for $\nC$. We present only
the important steps, the full derivation from first principles being described in the Supplemental Document, Sec.~II.

\subsection{Disorder model}

The real samples are composed of packed KDP crystal grains of different sizes and shapes. Thus the most relevant and
simple disorder model consists in a fluctuating continuous and real (no absorption) permittivity
$\epsilon(\bm{r})$. The disorder microstructure is then characterized by a spatial correlation function chosen to be Gaussian,
in the form
\begin{multline}\label{disorder_correlation}
   C_{\epsilon}(|\bm{r}-\bm{r}'|,\omega)=\bra\delta\epsilon(\bm{r},\omega)\delta\epsilon(\bm{r}',\omega)\ket
\\
      =|\Delta\epsilon(\omega)|^2\exp\left[-\frac{|\bm{r}-\bm{r}'|^2}{2\ell^2}\right].
\end{multline}
In this equation, $\delta\epsilon(\bm{r},\omega)=\epsilon(\bm{r},\omega)-\bra\epsilon(\bm{r},\omega)\ket$ is the
fluctuating part of the permittivity, $|\Delta\epsilon(\omega)|^2$ is the amplitude of the correlation and $\ell$ is the
correlation length. $|\Delta\epsilon(\omega)|^2$ depends on frequency since the permittivity $\epsilon$ is
dispersive. However, $\ell$ involves only the geometrical structure of the disorder and thus does not depend on
frequency. The $\chi^{(2)}$ nonlinearity is supposed to be correlated in a similar way.

\subsection{Linear regime}

We first consider the linear regime ($\lambda=\SI{1064}{nm}$) correspnding to propagation at the fundamental frequency
$\omega$. We use an approach similar to that in Ref.~\cite{PIERRAT-2008-1} developed in the context of the
Diffusing-Wave Spectroscopy (DWS).  The most important point concerns the selections of the scattering paths followed by
the field $E$ and its complex conjugate $\mE^*$ that dominate in the expression of the correlation function $\nC$. In a
dilute medium such that $k_0\ell_s\gg 1$, the leading contribution corresponds to $E$ and $\mE^*$ following the same
scattering sequences.  These sequences can be represented by the diagram
\begin{equation}\label{ladder}
   \begin{ddiag}{32}
      \rput(0,3){$E(\bm{r},\omega)$}
      \rput(0,-3){$\mE^*(\bm{r},\omega)$}
      \ggmoy{5}{10}{-3}
      \ggmoy{5}{10}{3}
      \ggmoy{10}{16}{-3}
      \ggmoy{10}{16}{3}
      \ggmoy{16}{22}{-3}
      \ggmoy{16}{22}{3}
      \eemoy{22}{27}{-3}
      \eemoy{22}{27}{3}
      \ccorreldeuxc{10}{3}{10}{-3}
      \ccorreldeuxc{16}{3}{16}{-3}
      \ccorreldeuxc{22}{3}{22}{-3}
      \pparticule{10}{-3}
      \pparticule{10}{3}
      \pparticule{16}{-3}
      \pparticule{16}{3}
      \pparticule{22}{-3}
      \pparticule{22}{3}
      \rput(30,3){$E_0$}
      \rput(30,-3){$E_0^*$}
   \end{ddiag}
\end{equation}
with an arbitrary number of scattering events \cite{kupri2017}. In these so-called ladder diagrams, the circles represent the
scattering events, the thick solid lines correspond to the Green functions modelling the field propagation between scattering
events and the thick dashed lines denote the incident field. The upper (bottom) line describes the propagation of $E$
($\mE^*$) respectively and the thin dashed vertical lines represent the disorder correlation $C$. It is important to note that in the model of continuous disorder, the circles do not represent real scatterers (grains) but scattering events connected by the correlation function $C$. The width $\ell$ of the correlation
function $C$ is however on the order of the grain size. The ladder shape of this dominant diagram implies that
there is always constructive interference between the field $E$ and its complex conjugate 
$\mE^*$. Thus the problem of computation of $\nC$ reduces to the problem of solving a radiative transport equation~\cite{RYTOV-1989}
\begin{multline}\label{linear_rte}
   \left[\bm{u}\cdot\bm{\nabla}_{\bm{r}}+\frac{1}{\ell_s(\omega)}\right]\mI(\bm{r},\bm{u},\omega)
\\
      =\frac{1}{\ell_s(\omega)}\int p(\bm{u},\bm{u}',\omega)g(\bm{r},\bm{u},\bm{u}',\omega)\mI(\bm{r},\bm{u}',\omega)\ud\bm{u}'.
\end{multline}
where $\mI(\bm{r},\bm{u},\omega)$ is the specific intensity, that can be seen as the radiative flux at position $\bm{r}$, in direction
$\bm{u}$ and at frequency $\omega$. More precisely, it can be shown from first principles that it is given by the Wigner
transform of the field. In our context (scatterer displacements), it reads
\begin{multline}\label{specific_intensity}
   \delta(k-k_0)\mI(\bm{r},\bm{u},\omega)=\int \bra E\left(\bm{r}+\frac{\bm{s}}{2},\omega\right)
                                                \mE^*\left(\bm{r}-\frac{\bm{s}}{2},\omega\right)\ket
\\\times
                                                e^{-ik\bm{u}\cdot\bm{s}}\ud\bm{s}.
\end{multline}
Thus, solving for $\mI$ gives a direct access to the field-field correlation function $\nC$. Equation~(\ref{linear_rte}) is very
similar to the standard Radiative Transfer Equation (RTE)~\cite{CHANDRASEKHAR-1950} except that it includes an
additional function $g(\bm{r},\bm{u},\bm{u}',\omega)$ that represents the decorrelation of the field at each scattering
event due to the motion of scatterers. It is given by
\begin{equation}\label{decorrel_function}
   g(\bm{r},\bm{u},\bm{u}',\omega)=\int e^{-ik_0(\bm{u}-\bm{u}')\cdot\bm{\Delta}}f(\bm{r},\bm{\Delta})\ud\bm{\Delta}
\end{equation}
where $\bm{u}$ and $\bm{u}'$ are unit vectors representing the scattered and incoming directions for a given scattering
process. $f(\bm{r},\bm{\Delta})$ is the probability density to have a displacement $\bm{\Delta}$ of a scatterer at the
position $\bm{r}$. The position dependence is required since this displacement is induced by the radiation pressure
that can be heterogeneous inside the medium (in particular at small depths). Eq.~(\ref{decorrel_function}) can be interpreted as follows: the
decorrelation is due to the phase shift (Doppler shift) averaged over all accessible displacements for a scatterer.
As a simple model, we consider that the amplitude of the displacement is proportional to the specific intensity which leads
to
\begin{multline}\label{displacement_probability}
   f\left(\bm{r},\bm{\Delta}\right)
      =\delta\left[\Delta-\beta I\left(\bm{r},\frac{\bm{\Delta}}{\Delta},\omega\right)\right]
\\\times
         \left[\beta^2\int I\left(\bm{r},\bm{u},\omega\right)^2\ud{\bm{u}}\right]^{-1}
\end{multline}
where $\beta$ is factor taking into account the link between the displacement and the value of the specific intensity.
In the following $\beta$ will be considered as a scaling parameter. In
Eq.~(\ref{displacement_probability}), $I\left(\bm{r},\bm{u},\omega\right)$ is the specific intensity without any
displacements. Finally $p(\bm{u},\bm{u}',\omega)$ is the phase function representing the part of energy incident from
direction $\bm{u}'$ and scattered into direction $\bm{u}$. For the Gaussian disorder considered here, it is given by
\begin{equation}
   p(\bm{u},\bm{u}',\omega)\propto\np(k_0|\bm{u}-\bm{u}'|)
   \quad\text{where}\quad
   \np(q)=\exp\left[-\frac{q^2\ell^2}{2}\right]
\end{equation}
and normalized such that $\int p(\bm{u},\bm{u}',\omega)\ud\bm{u}'=1$. Equation~(\ref{linear_rte}) can easily be interpreted using a 
random walk approach. Indeed, light undergoes a random walk whose average step is given by the scattering mean-free path 
$\ell_s(\omega)$ and whose angular distribution at each scattering event is given by the phase function $p(\bm{u},\bm{u}',\omega)$.
A phase shift is introduced between the fields at each scattering event due to the displacement of the scatterers as described by 
the function $g(\bm{r},\bm{u},\bm{u}',\omega)$.

\subsection{Second harmonic regime}

We now address the question of the generation and propagation of the second harmonic light. As it is usually done in homogeneous
materials, we use a perturbative approach in order to compute the field at $2\omega$ from the field at $\omega$. The full process
can be broken down into three steps. First the linear field at $\omega$ propagates inside the material. Second, it is converted
to second harmonic on an arbitrary scatterer. And finally, this process is followed by the propagation of the second
harmonic field. The same sequence of processes also applies to the complex conjugate of the field.  From this sequence, the
most important point is still here the identification of the leading diagram taking into account the non-linearity. It
is given by
\begin{equation}\label{non_linear_ladder}
   \begin{ddddiag}{60}
      \rput(0,3){$E(\bm{r},2\omega)$}
      \rput(0,-3){$\mE^*(\bm{r},2\omega)$}
      \ggmoy{5}{11}{-3}
      \ggmoy{11}{17}{-3}
      \ggmoy{17}{23}{-3}
      \ggmoy{23}{29}{-3}
      \ggmoy{29}{35}{-3}
      \ggmoy{35}{41}{-3}
      \ggmoy{41}{47}{-3}
      \eemoy{47}{53}{-3}
      \ggmoy{5}{11}{3}
      \ggmoy{11}{17}{3}
      \ggmoy{17}{29}{3}
      \ggmoy{29}{35}{3}
      \ggmoy{35}{41}{3}
      \ggmoy{41}{47}{3}
      \eemoy{47}{53}{3}
      \ccorreldeuxc{11}{-3}{11}{3}
      \ccorreldeuxc{17}{-3}{17}{3}
      \ccorreldeuxc{23}{-3}{23}{3}
      \ccorreldeuxc{29}{-3}{29}{3}
      \ccorreldeuxc{35}{-3}{35}{3}
      \ccorreldeuxc{41}{-3}{41}{3}
      \ccorreldeuxc{47}{-3}{47}{3}
      \pparticule{11}{-3}
      \pparticule{17}{-3}
      \pparticule{23}{-3}
      \pparticule{35}{-3}
      \pparticule{41}{-3}
      \pparticule{47}{-3}
      \pparticule{11}{3}
      \pparticule{17}{3}
      \pparticule{23}{3}
      \pparticule{35}{3}
      \pparticule{41}{3}
      \pparticule{47}{3}
      \gggmoy{29}{37}{3}{9}
      \ggmoy{37}{43}{9}
      \ggmoy{43}{49}{9}
      \eemoy{49}{55}{9}
      \ccorreldeuxc{37}{-9}{37}{-4}
      \ccorreldeuxc{37}{-2}{37}{2}
      \ccorreldeuxc{37}{4}{37}{9}
      \ccorreldeuxc{43}{-9}{43}{-4}
      \ccorreldeuxc{43}{-2}{43}{2}
      \ccorreldeuxc{43}{4}{43}{9}
      \ccorreldeuxc{49}{-9}{49}{-4}
      \ccorreldeuxc{49}{-2}{49}{2}
      \ccorreldeuxc{49}{4}{49}{9}
      \pparticule{37}{9}
      \pparticule{43}{9}
      \pparticule{49}{9}
      \gggmoy{29}{37}{-3}{-9}
      \ggmoy{37}{43}{-9}
      \ggmoy{43}{49}{-9}
      \eemoy{49}{55}{-9}
      \pparticule{37}{-9}
      \pparticule{43}{-9}
      \pparticule{49}{-9}
      \nnonlineaire{29}{-3}
      \nnonlineaire{29}{3}
      \rput(58,9){$E_0$}
      \rput(58,3){$E_0$}
      \rput(58,-3){$E_0^*$}
      \rput(58,-9){$E_0^*$}
   \end{ddddiag}
\end{equation}
where the squares represent the second harmonic generation process. We could assume the non-linear processes for $E$ and
$\mE^*$ to occur at two positions with an arbitrary distance between them. However, this would lead to the propagation
of the correlations $\bra E(\bm{r},\omega)\mE^*(\bm{r},2\omega)\ket$ or $\bra E(\bm{r},2\omega)\mE^*(\bm{r},\omega)\ket$
which are supposed to vanish since they involve fields at two different frequencies. The generations of the second harmonic
sources for $E$ and $\mE^*$ are then confined in a small volume with typical size $\ell$. The relevance of the dominant
diagram responsible for the second harmonic correlation has to be carefully checked. For that purpose, we have performed
ab initio numerical simulations that are presented in the Supplemental Document, Sec.~III. Finally, the diagram of Eq.~(\ref{non_linear_ladder}) can be
interpreted the following way: the right part represents the propagation of two specific intensities at frequency
$\omega$ obeying Eq.~(\ref{linear_rte}) and the left part represents the propagation of the specific intensity
at frequency $2\omega$. It is given by the following non-linear RTE
\begin{multline}\label{non_linear_rte}
   \left[\bm{u}\cdot\bm{\nabla}_{\bm{r}}+\frac{1}{\ell_s(2\omega)}\right]\mI(\bm{r},\bm{u},2\omega)
\\
      =\frac{1}{\ell_s(2\omega)}\int p(\bm{u},\bm{u}',2\omega)g(\bm{r},\bm{u},\bm{u}',2\omega)\mI(\bm{r},\bm{u}',2\omega)\ud\bm{u}'
\\
      +\alpha\iint \pSHG(\bm{u},\bm{u}',\bm{u}'',\omega)\gSHG(\bm{r},\bm{u},\bm{u}',\bm{u}'',\omega)
\\
         \times\mI(\bm{r},\bm{u}',\omega)\mI(\bm{r},\bm{u}'',\omega)\ud\bm{u}'\ud\bm{u}''.
\end{multline}
This equation is the main theoretical result of this work. It shows that the second harmonic specific intensity follows a similar
transport equation as the fundamental intensity [Eq.~(\ref{linear_rte})], but with a source term describing the non-linear conversion
process. Its physical interpretation is very simple: Light propagates first at frequency $\omega$
which is described by the specific intensity
$\mI(\bm{r},\bm{u}'',\omega)$ solution to Eq.~(\ref{linear_rte}). Then a SHG process occurs which creates a source at
frequency $2\omega$, the amplitude of which is given by the product of two specific intensities at $\omega$. Finally,
propagation at $2\omega$ is described by the specific intensity $\mI(\bm{r},\bm{u}'',2\omega)$ that follows 
Eq.~(\ref{non_linear_rte}). In this expression, $\alpha$ is a factor that takes into account all constants involved in
the SHG process such as $\chi^{(2)}$. $\pSHG(\bm{u},\bm{u}',\bm{u}'',\omega)$ is the SHG phase function. It involves
three units vectors. $\bm{u}'$ and $\bm{u}''$ corresponds to the incoming directions of the two specific
intensities at $\omega$ and $\bm{u}$ is the outgoing direction of the specific intensity at 2$\omega$. In the case of the correlated
disorder we consider here, we have
\begin{equation}
   \pSHG(\bm{u},\bm{u}',\bm{u}'',\omega)\propto\np(k_0|2\bm{u}-\bm{u}'-\bm{u}''|)
\end{equation}
with $\int \pSHG(\bm{u},\bm{u}',\bm{u}'',\omega)\ud\bm{u}'\ud\bm{u}''=1$. $\bm{u}$ appears with a factor two since it
corresponds to the non-linear specific intensity direction.
$\gSHG(\bm{u},\bm{u}',\bm{u}'',\omega)$ is the decorrelation function given by
\begin{equation}\label{non_linear_decorrel_function}
   \gSHG(\bm{r},\bm{u},\bm{u}',\bm{u}'',\omega)=\int
      e^{-ik_0(2\bm{u}-\bm{u}'-\bm{u}'')\cdot\bm{\Delta}}f(\bm{r},\bm{\Delta})\ud\bm{\Delta}.
\end{equation}
It still corresponds to the decorrelation induced by a Doppler shift involving three beams, \ie two incoming beams at frequency $\omega$ in directions $\bm{u}'$ and $\bm{u}''$ and one outgoing beam at frequency $2\omega$ in direction $\bm{u}$.

\subsection{Numerical simulations}

\begin{figure}[h]
   \centering
   \includegraphics[width=\linewidth]{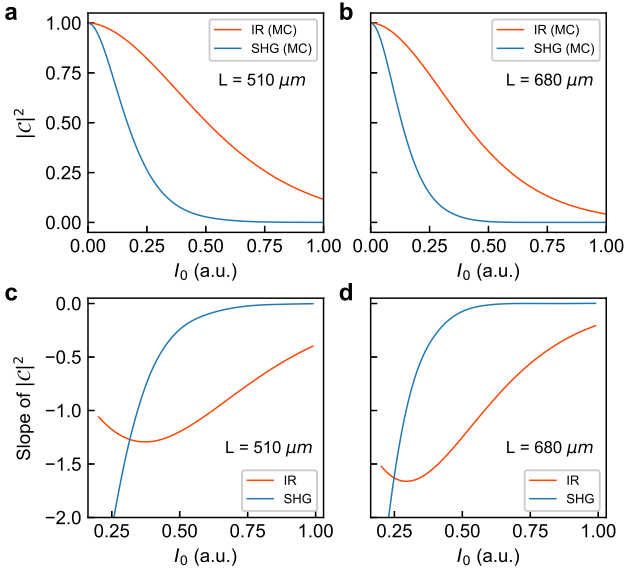}
   \caption{Monte Carlo (MC) simulated correlation functions $|\nC|^2$ for the linear (red solid line) and second
   harmonic (blue solid line) beams as a function of the incident intensity $I_0$ in arbitrary units. (a)
   $L=510$~$\mu$m, (b) $L=680$ $\mu$m. The range of $I_0$ is chosen such that the extreme values of the linear
   correlation $|\nC(\omega)|^2$ are in agreement with the experimental values $\nC_{\text{expe}}(\omega)$ presented.
   Slopes of  $|\nC(\omega)|^2$ and $|\nC(2\omega)|^2$ calculated on each power value  for the two samples are shown in
   (c) $L=\SI{510}{\micro\meter}$ and (d) $L=\SI{680}{\micro\meter}$. A qualitative agreement is immediately seen with
   the experimental behavior in Fig.~\ref{fig:3}}
   \label{fig:results_theo}
\end{figure}

In order, to solve the system of Eqs.~(\ref{linear_rte}) and (\ref{non_linear_rte}), we have developed a Monte Carlo
scheme which can be seen as a random walk process inside the material~\cite{SIEGEL-1992}. Three Monte Carlo simulations
are performed in a slab geometry of thickness $L$ under plane-wave illumination at normal incidence. The first is used
to compute $I\left(\bm{r},\bm{u},\omega\right)$, the specific intensity in the absence of
displacement of the scatterers. This is required in order to compute the probability density $f(\bm{r},\bm{\Delta})$ to have a
displacement $\bm{\Delta}$ at position $\bm{r}$. The second Monte Carlo simulation is used to compute
$\mI\left(\bm{r},\bm{u},\omega\right)$, the specific intensity associated with the correlation function at $\omega$.
Finally, a last simulation is performed in order to compute $\mI\left(\bm{r},\bm{u},2\omega\right)$, the specific
intensity associated with the correlation function at $2\omega$. More precisely, the correlation functions are computed
from the energy density at the output interface in transmission, \ie
\begin{align}
   \nC(\omega) & =\frac{\int \mI\left(z=L,\bm{u},\omega\right)\ud\bm{u}}
      {\int I\left(z=L,\bm{u},\omega\right)\ud\bm{u}},
\\
   \nC(2\omega) & =\frac{\int \mI\left(z=L,\bm{u},2\omega\right)\ud\bm{u}}
      {\int I\left(z=L,\bm{u},2\omega\right)\ud\bm{u}}.
\end{align}
This computations are performed for different incident intensities $I_0$ (or different incident powers $\bra P\ket$) meaning
different probability density $f(\bm{r},\bm{\Delta})$ which correspond to different radiation pressures. Regarding the
numerical parameters, it is important to keep in mind that the KDP powder used in the experiment has crystal
grains of different sizes ranging from \SI{2}{\micro\meter} to \SI{8}{\micro\meter}. This makes difficult the choice of
the correlation length $\ell$. However we have tested several values showing that this is not a crucial parameter.
Since the particles are large compared to the wavelength, we have chosen $k_0\ell=3$ for the results presented
in Fig.~\ref{fig:results_theo}. This gives the anisotropy factors $g(\omega)=0.89$ and $g(2\omega)=0.97$. The thickness of
the medium $L$ as well as the transport mean-free paths $\ell_t=\ell_s/(1-g)$ take the values measured experimentally,
which gives $k_0L=4016$, $k_0\ell_t(\omega)=2079$ and $k_0\ell_t(2\omega)=998$.
This finally leads to the normalized scattering mean-free paths $k_0\ell_s(\omega)=229$ and $k_0\ell_s(2\omega)=30$.

\begin{figure*}[ht]
   \centering
   \includegraphics[width=\linewidth]{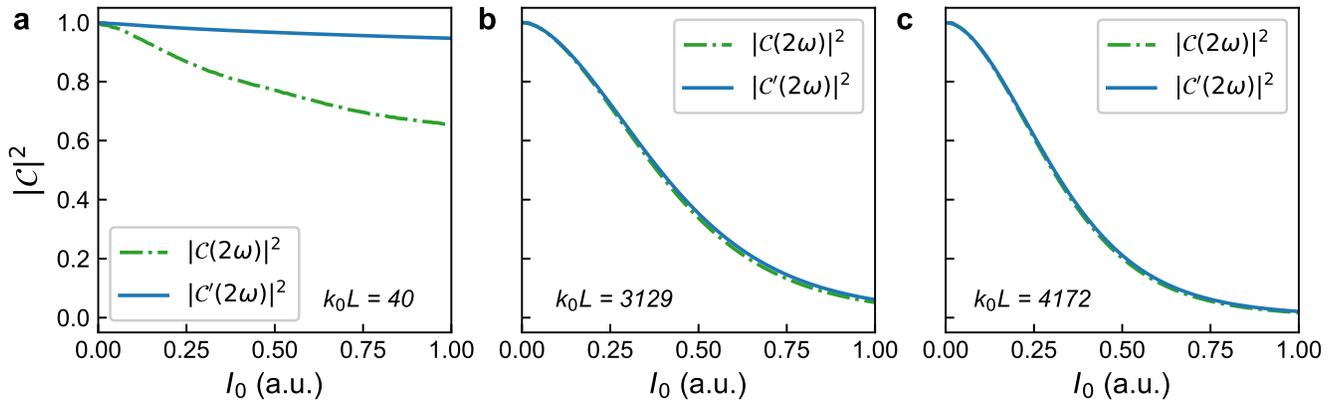}
   \caption{Monte Carlo (MC) simulated correlation functions $|\nC|^2$ for the second harmonic speckles taking into account all decorrelation effects
   (\ie effect of $g$ and $\gSHG$, green dashed line) and only the decorrelation occurring during beam propagation (\ie
   effect of $g$, blue solid line). The computations are performed for  (a) $k_{0}L=40$, (b) $k_{0}L = 3129$ and (c) $k_{0}L=4172$. The other parameters are the same as in Fig.~\ref{fig:results_theo}.}
   \label{results_g_g_shg}
\end{figure*}

We can observe that the results shown in Fig.~\ref{fig:3} are reproduced qualitatively. The second-harmonic beam
decorrelates faster than that  of the fundamental frequency. The slopes of the decorrelation curves also reproduce the
same trends seen in experiments, for most part of the input power range. The IR light shows a reducing slope followed by
a rise at a certain pump power, while the frequency-doubled light shows a steady rise followed by a saturation region.
Barring a minor difference at low powers, the experimental data exhibit the same behavior. The crossing of the two slope
curves also happens at a lower pump power in the thicker sample, as seen in the experiments. The agreement with the
experiments is very clear qualitatively, but is not quantitative. The main reason should probably be investigated in the
dependence of the probability density $f$ on the specific intensity $I$. Building this relationship is not a trivial
task and is out of the scope of the present work. A second potential effect that has been neglected so far, is the role
of the refractive index mismatch at the interfaces of the slab. In the Monte Carlo simulation we have verified that this
does not change substantially the results up to a refractive index $n=2$.

The faster decorrelation of the second harmonic speckle can be explained through two different mechanisms: The
first one corresponds to the decorrelation when the second harmonic light is generated which is represented by
$\gSHG(\bm{r},\bm{u},\bm{u}',\bm{u}'',\omega)$. Its dependence on the three different directions through the relation
$2\bm{u}-\bm{u}'-\bm{u}''$ favors a faster decorrelation. The second mechanism is due to the propagation of the
second harmonic field. The factor of two in $g(\bm{r},\bm{u},\bm{u}',2\omega)$ also makes the correlation vanish
faster than for the linear beam. For a small optical thickness, both effects play a role and have to be taken into
account properly. This comes from the fact that photons experience few scattering events before escaping the medium and
thus the decorrelations due to $\gSHG$ and to $g$ are of the same order of magnitude. On the other hand, for large
optical thicknesses, many scattering events are involved and the contribution of $\gSHG$ is negligible compared to that
of $g$. Figure~\ref{results_g_g_shg} illustrate this statement using the Monte Carlo simulations. This simple conclusion
will also be easily observed in the more simple case of statistically homogeneous and isotropic displacements discussed
in the next section.

\subsection{Statistically homogeneous and isotropic displacements}

Beyond the effect of radiation pressure that has been examined in this study, it is
also interesting to consider a displacement probability for the scatterers that is homogeneous and isotropic, \ie
$f(\bm{r},\bm{\Delta})=f(\Delta)/(4\pi\Delta^2)$. Indeed, considering large medium thicknesses compared to the transport
mean-free paths, \ie $L\gg\{\ell_t(\omega),\ell_t(2\omega)\}$, we can derive diffusion equations for the linear and
second harmonic correlation functions. The derivation is detailed in the Supplemental Document, Sec.~IV and
leads to analytical expressions given by
\begin{align}
   \nC(\omega) & = \frac{\kappa(\omega) L}{\sinh[\kappa(\omega) L]},
\\\nonumber
   \nC(2\omega) & = \frac{6\nD(\omega)}{\kappa(2\omega)L\sinh[\kappa(\omega)L]^2\sinh[\kappa(2\omega)L]}
\\ & \hspace{-6ex}\times
      \frac{\kappa(2\omega)^2\left\{1-\cosh[2\kappa(\omega)L]\right\}
         -4\kappa(\omega)^2\left\{1-\cosh[\kappa(2\omega)L]\right\}}{\kappa(2\omega)^2-4\kappa(\omega)^2}
\end{align}
where
\begin{align}
   \kappa & =\sqrt{\frac{3}{\mell_t\mell_a}},
               \quad \mell_t=\frac{\ell_s}{1-\mg},
\\
   \frac{1}{\mell_a} & = \frac{1}{\ell_s}\left[1-\int p(\bm{u},\bm{u}')g(\bm{u},\bm{u}')\ud\bm{u}'\right],
\\
   \text{and } \mg & = \frac{\int p(\bm{u},\bm{u}')g(\bm{u},\bm{u}')\bm{u}\cdot\bm{u}'\ud\bm{u}'}
                         {\int p(\bm{u},\bm{u}')g(\bm{u},\bm{u}')\ud\bm{u}'},
\end{align}
all these four quantities being defined at $\omega$ and $2\omega$. We also have
\begin{equation}
   \nD(\omega)=\frac{\int
       \pSHG(\bm{u},\bm{u}',\bm{u}'',\omega)\gSHG(\bm{u},\bm{u}',\bm{u}'',\omega)\ud\bm{u}\ud\bm{u}'\ud\bm{u}''
      }{\int \pSHG(\bm{u},\bm{u}',\bm{u}'',\omega)\ud\bm{u}\ud\bm{u}'\ud\bm{u}''}.
\end{equation}
We clearly see from these expressions that the effect of the decorrelation during the propagation of the waves at
$\omega$ or at $2\omega$ can be seen as an absorption effect and are encoded in the $\kappa$ functions. The decorrelation
process taking place during the generation of the second harmonic light is encoded in the $\nD$ function. Finally these
analytical expressions can be simplified in the even more particular case of isotropic scattering such that
$p(\bm{u},\bm{u}')=1/(4\pi)$ and $\pSHG(\bm{u},\bm{u}',\bm{u}'',\omega)=1/(16\pi^2)$ and of a constant displacement
amplitude $d$ such that $k_0d\ll 1$ and $f(\Delta)=\delta(\Delta-d)$. This gives
\begin{align}
   \kappa(\omega)L & =b(\omega)k_0d,
   \quad
   \kappa(2\omega)L=2b(2\omega)k_0d
\\\nonumber
   \text{and } \nD(\omega) & = \frac{1}{8\pi}\int_{\phi=0}^{2\pi}\int_{\mu=-1}^1\int_{\mu'=-1}^1
         \operatorname{sinc}\left[k_0d
            \vphantom{\sqrt{6-4\mu-4\mu'+2\mu\mu'+2\sqrt{1-\mu^2}\sqrt{1-\mu'^2}\cos\phi}}
         \right.
\\\nonumber & \hspace{-5ex}\left.\times
         \sqrt{6-4\mu-4\mu'+2\mu\mu'+2\sqrt{1-\mu^2}\sqrt{1-\mu'^2}\cos\phi}\right]
\\ & \times
            \ud\mu\ud\mu'\ud\phi
\end{align}
where $b=L/\ell_s$ is the optical thickness. These last expressions are very useful to get more insights on the 
decorrelation effects encoded in functions $\kappa(\omega)$, $\kappa(2\omega)$ and $\nD(\omega)$. In particular, as
already noticed in the numerical simulations, we clearly see that the decorrelation during propagation is stronger when the optical thicknesses $b(\omega)$ and $b(2\omega)$ increase which reduces the effect of $\nD(\omega)$.
In the diffusive regime considered here, $\nD(\omega)$ can thus be replaced by its limit when $k_0d\to 0$, \ie 
$\nD(\omega)\sim 1$.

\section{Discussion and Conclusion}

In summary, we have experimentally investigated the decorrelation of speckle patterns with increasing pump power in a
second-order nonlinear disordered medium. Simultaneous speckle measurements at the fundamental and second harmonic
wavelengths reveal a varying rate of decorrelation under the same incident power. The decorrelation arises from the
microscopic displacements in the disorder configuration induced by the radiation pressure produced by the pump beam. In
addition, the second harmonic correlation decreases faster than the fundamental. We laid the foundations of a
theoretical model that accurately describes the synergy of second-order nonlinearity and light diffusion. The model
demarcates the contribution of two components in the decorrelation, namely, one arising from the generation of
second-harmonic light, and the other arising from the propagation thereof. For the samples and input powers employed in
our experiments, the former seems to be the stronger contributor. Wider investigations of the model show that the
relative strengths of the two components depend upon the degree of disorder. Towards the differences in the experimental
and computed results, we have discussed qualitatively the origins as follows. The actual displacement at a location
$\bm{r}$ is dependent on the specific intensity at that location, and the size and shape of the particle at that
location. This is too intricate a parameter to calculate, and we did not venture into it. In the theory, the sample is
homogeneously disordered, and particle size is not a parameter in computing the displacement under radiation pressure.
At a future stage, a distribution in the displacements may be invoked in the theory. We believe these unavoidable
differences in the experimental samples and theoretical assumptions limit the agreement in the respective results. This
study shades light on the subtle mechanism of non-linear conversion in disordered media, with expected outcomes in
fundamental studies in mesoscopic wave transport, as well as the design of efficient materials for non-linear generation
of light.

\section*{Funding Information} 

Department of Atomic Energy, Government  of  India for funding for the project identification No. RTI4002 under the DAE
OM No 1303/1/2020/R\&D-II/DAE/5567 dated 20.8.2020; Swarnajayanti Fellowship, Department of Science and Technology,
Ministry of Science and Technology, India.

This work has received support under the program ``Investissements d’Avenir'' launched by the French Government.

\section*{Acknowledgment}

R.S. and S.M. acknowledge the support from Sandip Mondal and N. Sreeman Kumar during the experiment.

\section*{Disclosures}

The authors declare no conflicts of interest.

\bibliography{shg}

\end{document}


\title{Speckle decorrelation in fundamental and second-harmonic light scattered from nonlinear disorder: supplemental document}

\author{Rabisankar Samanta}
\affiliation{Nano-optics and Mesoscopic Optics Laboratory, Tata Institute of Fundamental Research, 1 Homi Bhabha Road, Mumbai, 400 005, India}
\author{Romain Pierrat}
\email{romain.pierrat@espci.psl.eu}
\affiliation{Institut Langevin, ESPCI Paris, PSL University, CNRS, 1 rue Jussieu, 75005 Paris, France}
\author{Rémi Carminati}
\affiliation{Institut Langevin, ESPCI Paris, PSL University, CNRS, 1 rue Jussieu, 75005 Paris, France}
\affiliation{Institut d’Optique Graduate School, Université Paris-Saclay, F-91127 Palaiseau, France}
\author{Sushil Mujumdar}

\email{mujumdar@tifr.res.in}
\affiliation{Nano-optics and Mesoscopic Optics Laboratory, Tata Institute of Fundamental Research, 1 Homi Bhabha Road, Mumbai, 400 005, India}

\date{\today}

\maketitle

\tableofcontents
 
\section{A different experiment to validate particle movement due to laser radiation pressure}

\begin{figure*}
   \begin{minipage}{0.42\linewidth}
      \centering
      \includegraphics[width=\linewidth]{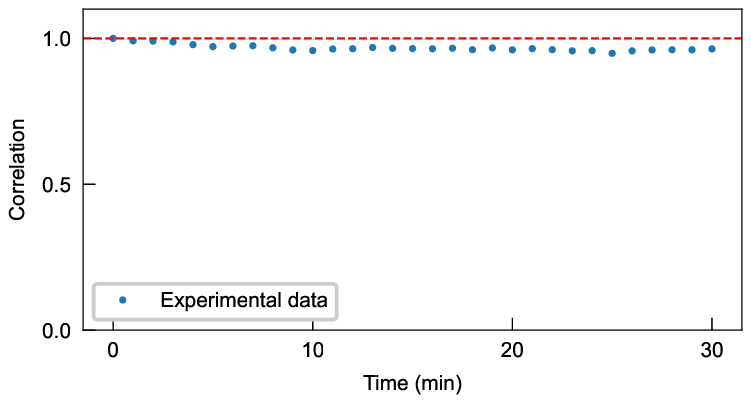}
      \caption{Speckle correlation showing the high stability of the HeNe laser.}
      \label{fig:HeNe}
   \end{minipage}
   \hfill
   \begin{minipage}{0.52\linewidth}
      \centering
      \includegraphics[width=\linewidth]{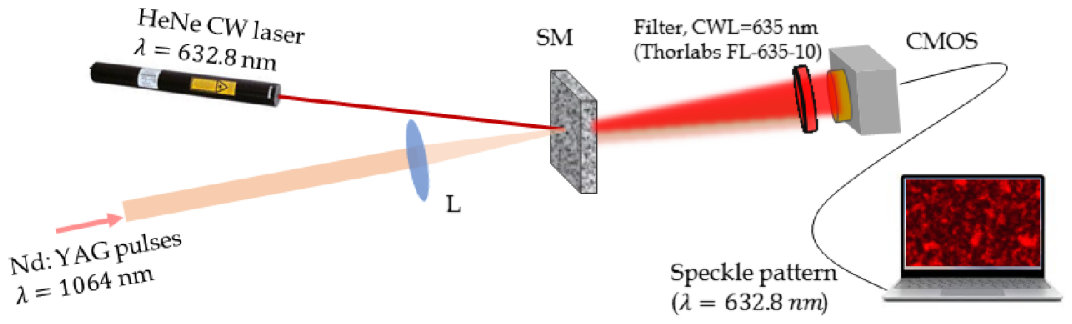}
      \caption{Experimental setup to investigate particles’ movements due to high input laser power. L: Lens, SM: Scattering medium.}
      \label{fig:HeNe_speckle_setup}
   \end{minipage}
\end{figure*}

We have carried out experiments to demonstrate the scatterer displacements are due to radiation pressure. Specifically, a stabilized frequency, low power CW laser (HeNe, $\lambda=\SI{632.8}{nm}$) was made incident onto the sample simultaneously with the pump. The phase stability of the HeNe laser was confirmed by performing a simple experiment. First, we focused the HeNe laser on a ground glass diffuser and subsequently captured the transmitted speckle pattern on a CMOS camera over \SI{30}{min} at an interval of \SI{1}{min}. In Fig.~\ref{fig:HeNe}, we plot the speckle correlation calculated with respect to the initial speckle pattern. We observed that the correlation coefficient always stays close to the ideal value 1 (indicated by red dashed line) which proves our HeNe laser was to be highly stable.

\begin{figure*}
   \centering
   \includegraphics[width=0.6\linewidth]{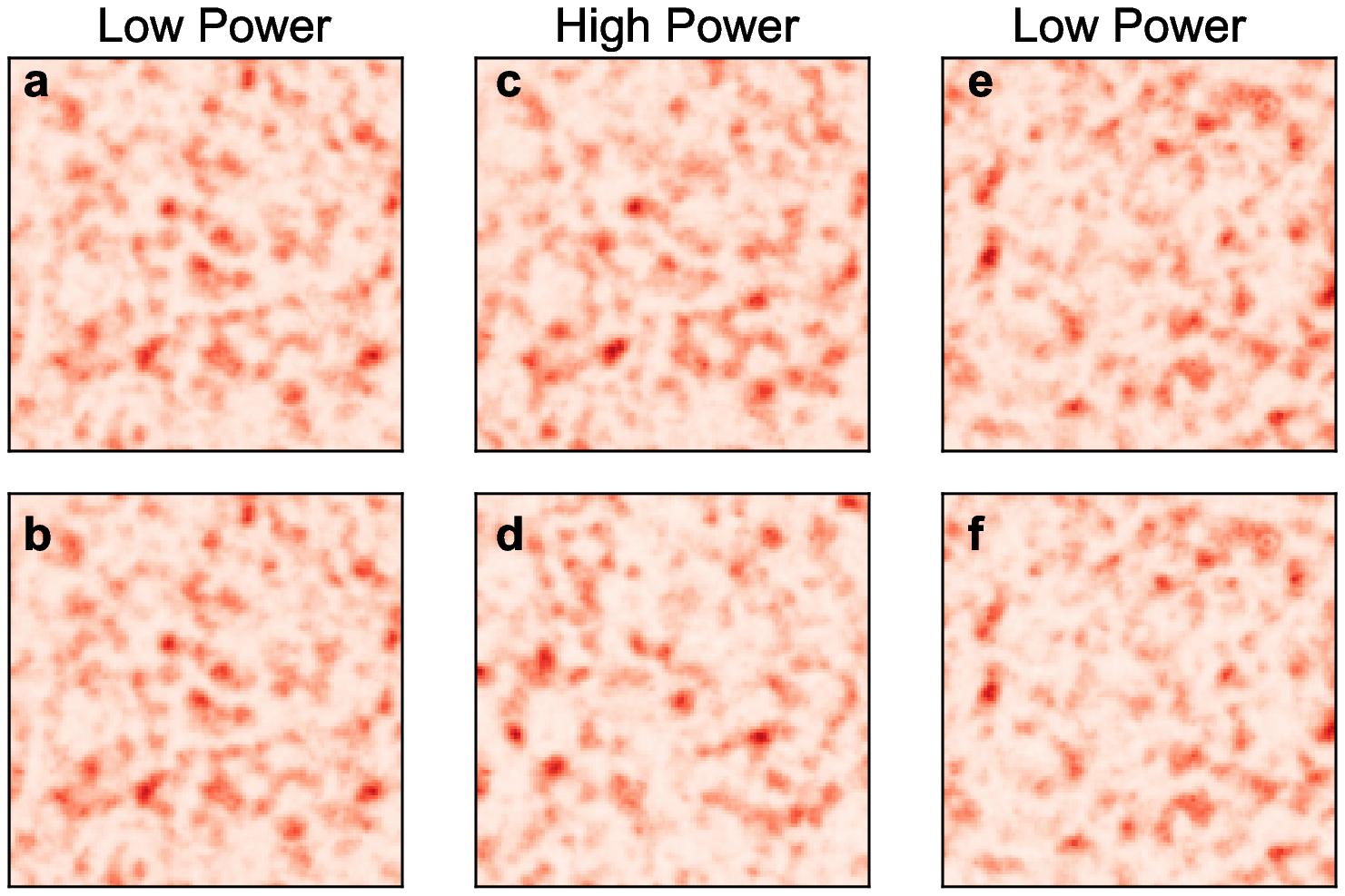}
   \caption{Speckle patterns of red light at different scenarios. (a, b) Beginning of the experiment, power of the input pulsed laser: low . (c, d) power of the input pulsed laser: high. (e, f) Ending of the experiment, power of the input pulsed laser: low.}
   \label{fig:HeNe_speckle}
\end{figure*}

It is well known that the speckle pattern produced by a scattering medium remains static if the scatterers of the medium do not move. A schematic of the experimental setup is depicted in Fig. \ref{fig:HeNe_speckle_setup}. Nd:YAG laser pulses (pulse width \SI{35}{ps}, repetition rate \SI{1}{Hz}) with fundamental wavelength of \SI{1064}{nm} were falling normally on the sample. A CW phase stabilized HeNe laser with wavelength of \SI{632.8}{nm} was also made incident on the sample at the same position. Here, it should be noted that the sample behaves as a linear medium for the laser coming from the HeNe whereas Nd:YAG pulses, at a sufficient input power, can generate nonlinear photons. As a result, in transmission, we could retrieve three speckle patterns, namely, red ($\lambda=\SI{632.8}{nm}$), IR ($\lambda=\SI{1064}{nm}$) and SH ($\lambda=\SI{532}{nm}$). For the current purpose, the speckle dynamics of the red light was monitored. A laser line filter (Thorlabs, FL-635-10) with a center wavelength of \SI{635}{nm} was placed in front of a CMOS camera (Thorlabs). We captured the speckle patterns of red light while increasing the Nd:YAG laser power from very low to high and again high to low. Speckles are reported in Fig.~\ref{fig:HeNe_speckle}\,(a-f).

The first two images [(a) and (b)] correspond to consecutive Nd:YAG laser shots where the input power is very low. Clearly, the speckle patterns are almost the same over these shots. Next, we increased the input power of the Nd:YAG laser such that the second harmonic was efficiently generated. Images (c) and (d) are consecutive speckle images corresponding to this input power, and they are seen to be obviously different. This shows that the configuration of the sample was changing from shot to shot at high power. Next, we again decreased the Nd:YAG power to the minimum power and grabbed images (e) and (f) on consecutive laser pulses. Again, the speckle becomes stable when the input power reaches a very low value, since the sample is stable. Further, it is also observed that the initial [(a) and (b)] and final [(e) and (f)] red speckle patterns, although at the same low power, are completely different, which indicates that the particle configuration inside the sample has changed over the experiment.  All of this clearly indicates the scatterer displacements due to the high input power.\par

This experiment allows us to rule out thermal effects. Thermal effects tend to work on longer timescales than impact by the radiation pressure. In the speckle pattern generated by the HeNe laser, we observed that the speckle was instantaneously reorganized when the pulse hit the sample, and immediately thereafter, the pattern was stable. There was no gradual and continuous change in the speckle for a finite time after the pulse. This motivates us to claim that, although thermal effects cannot be ruled out completely, the major contribution to the decorrelation arises from displacement due to pulse impact. Furthermore, the laser pulses were shot at too low a repetition rate to expect a cumulative thermal effect.

\section{Transport model derivation}

This section is dedicated to the derivation of the transport model used to estimate the correlation function $\nC$ in
the linear as well as in the second harmonic regime. This derivation is largely inspired to the standard multiple
scattering theory whose details can be found in several references (see \eg Refs.~\onlinecite{RYTOV-1989,APRESYAN-1996,ROSSUM-1999}). The
generalization presented here allows to take into account (1) scatterer displacements (2) second harmonic generation.
Regarding the first point, the reader can also refer to Ref.~\onlinecite{PIERRAT-2008-1} that presents a similar derivation.
Regarding the second point, a similar approach has been used in the context of photon echoes described
in Ref.~\onlinecite{PIERRAT-2018}.

\subsection{Disorder model and scattering potential}

We consider a continuous disorder model described by a fluctuating permittivity $\epsilon(\bm{r},\omega)$ and a
fluctuating second order nonlinear susceptibility $\chi(\bm{r})$. We choose to characterize their statistical properties by a
Gaussian correlation function identical for both $\epsilon$ and $\chi$ since the disorder at the origin of the
scattering and of second harmonic generation are the same. This gives
\begin{align}\label{epsilon_correlation}
   C_{\epsilon}(\bm{r},\bm{r}',\omega) & =\bra\delta\epsilon(\bm{r},\omega)\delta\epsilon(\bm{r}',\omega)\ket
      =|\Delta\epsilon(\omega)|^2C(|\bm{r}-\bm{r}'|),
\\\label{chi_correlation}
   C_{\chi}(\bm{r},\bm{r}') & =\bra\delta\chi(\bm{r})\delta\chi(\bm{r}')\ket
      =|\Delta\chi|^2C(|\bm{r}-\bm{r}'|),
\\\label{gaussian}
   \text{with } C(|\bm{r}-\bm{r}'|) & =\exp\left[-\frac{|\bm{r}-\bm{r}'|^2}{2\ell^2}\right]
\end{align}
and where $|\Delta\epsilon(\omega)|^2$ and $|\Delta\chi|^2$ are the amplitudes and $\ell$ the correlation length.
$\delta\epsilon(\bm{r},\omega)=\epsilon(\bm{r},\omega)-\bra\epsilon(\omega)\ket$ is the fluctuating part of the
permittivity. Similarly $\delta\chi(\bm{r})=\chi(\bm{r})-\bra\chi\ket$ is the fluctuating part of the second order
susceptibility.  The disorder correlation function depends only on $|\bm{r}-\bm{r}'|$ because we consider that the
disorder is statistically isotropic and homogeneous. From this model, we first define the scattering potential which is
at the root of the multiple scattering theory. It is given by
\begin{equation}
   V(\bm{r},\omega)=k_0^2\left[\epsilon(\bm{r},\omega)-\epsilon_b(\omega)\right].
\end{equation}
$\epsilon_b$ is the background permittivity corresponding to the reference non-scattering medium. It is thus
homogeneous. We now consider that there are scatterer displacements for example under the action of the radiation pressure
but it can be a completely different physical process. After displacements, the new potential is given by
\begin{equation}\label{new_potential}
   \mV(\bm{r},\omega)=V(\bm{r}-\bm{\Delta}(\bm{r}),\omega)
\end{equation}
where $\bm{\Delta}(\bm{r})$ is the displacement at position $\bm{r}$. In the following, we consider that this
displacement is constant over a length scale of the order of $\ell$. This allows to take the Fourier transform of
Eq.~(\ref{new_potential}) considering no position dependence in $\bm{\Delta}$, which gives
\begin{equation}\label{new_potential_fourier}
   \mV(\bm{k},\omega)=V(\bm{k},\omega)\exp\left[-i\bm{k}\cdot\bm{\Delta}(\bm{r})\right].
\end{equation}
In a similar way, we have
\begin{equation}\label{new_chi}
   \mchi(\bm{r})=\chi(\bm{r}-\bm{\Delta}(\bm{r}))
\end{equation}
which leads to
\begin{equation}\label{new_chi_fourier}
   \mchi(\bm{k})=\chi(\bm{k})\exp\left[-i\bm{k}\cdot\bm{\Delta}(\bm{r})\right].
\end{equation}

\subsection{Self-energy and intensity vertex}

We focus now on the computation of some important building blocks regarding wave propagation in complex media that
are the self-energy $\Sigma$ and the intensity vertex $\Gamma$. Similarly to the scattering potential, we show how their
expressions are modified in order to take into account the scatterer displacements.

We first consider the self-energy $\Sigma$ and more importantly its counterpart
denoted by $\mSigma$ when scatterer displacements are present. The self-energy is an important quantity entering the
Dyson equation that drives the evolution of the statistical average electric field propagating inside a strongly
scattering medium. It contains all possible scattering sequences that cannot be statistically factorized. In a dilute
medium where the scattering mean-free path is large compared to the wavelength, it can be limited to the first two
orders which writes
\begin{equation}\label{sigma}
   \Sigma(\bm{r},\bm{r}',\omega)
      =\bra V(\bm{r},\omega)\ket\delta(\bm{r}-\bm{r}')
         +\bra V(\bm{r},\omega)G_b(\bm{r}-\bm{r}',\omega)V(\bm{r}',\omega)\ket_c
\end{equation}
where $\bra\cdot\ket_c$ represents a statistical average restricted to the connected part, \ie $\bra VG_bV\ket_c=\bra
VG_bV\ket - \bra V\ket G_b\bra V\ket$. $G_b$ is the Green function in the reference medium, it describes the field
propagation between two consecutive scattering events on the potential $V$. For the case of scatterer displacements, we
have
\begin{equation}\label{new_sigma}
   \mSigma(\bm{r},\bm{r}',\omega)
      =\bra \mV(\bm{r},\omega)\ket\delta(\bm{r}-\bm{r}')
         +\bra \mV(\bm{r},\omega)G_b(\bm{r}-\bm{r}',\omega)\mV(\bm{r}',\omega)\ket_c.
\end{equation}
It is important to keep in mind that the statistical average performed here is not only an average over all possible
configurations of the disorder but is also an average over the displacements.  By Fourier transforming
Eqs.~(\ref{new_sigma}) and (\ref{sigma}) and using Eq.~(\ref{new_potential_fourier}), we obtain
\begin{equation}
   \mSigma(\bm{k},\bm{k}',\omega)
      =\Sigma(\bm{k},\bm{k}',\omega)
         \int\exp\left[-i(\bm{k}-\bm{k}')\cdot\bm{\Delta}\right]f(\bm{r},\bm{\Delta})\ud\bm{\Delta}
\end{equation}
where $f(\bm{r},\bm{\Delta})$ is the probability density to have a displacement $\bm{\Delta}$ at position $\bm{r}$.
From the statistical homogeneity and isotropy of the disorder, we get
\begin{equation}
   \Sigma(\bm{k},\bm{k}',\omega)=8\pi^3\delta(\bm{k}-\bm{k}')\breve{\Sigma}(k,\omega)
   \quad\text{and}\quad
   \mSigma(\bm{k},\bm{k}',\omega)=8\pi^3\delta(\bm{k}-\bm{k}')\breve{\mSigma}(k,\omega)
\end{equation}
which leads to
\begin{equation}
   \breve{\mSigma}(k,\omega)=\breve{\Sigma}(k,\omega)
\end{equation}
where $\breve{\Sigma}$ and $\breve{\mSigma}$ are the reduced self-energies. Since the extinction mean-free path is given
by the imaginary part of the reduced self-energy, we have
\begin{equation}
   \frac{1}{\ell_e(\omega)}=\frac{\im\breve{\Sigma}(k_0,\omega)}{k_0}
      =\frac{\im\breve{\mSigma}(k_0,\omega)}{k_0}=\frac{1}{\mell_e(\omega)}.
\end{equation}
By invoking the non-absorbing nature of the medium we finally get equality between the scattering mean-free paths with
and without scatterer displacements. Inserting the expression of the correlation function $C_{\epsilon}$ into
Eq.~(\ref{sigma}) leads to
\begin{equation}
   \frac{1}{\mell_s(\omega)}=\frac{1}{\ell_s(\omega)}=\frac{k_0^4|\Delta\epsilon(\omega)|^2}{16\pi^2}
      \int C(q)\ud\Omega
\end{equation}
where $q=2k_0\sin(\theta/2)$ is the modulus of the scattering vector and $\ud\Omega=\sin\theta\ud\theta\phi$ is the
elementary solid angle in standard spherical units. From Eq.~(\ref{gaussian}), we get
\begin{equation}
   C(q)=\int\exp\left[-\frac{\bm{R}^2}{2\ell^2}-i\bm{q}\cdot\bm{R}\right]\ud\bm{R}
      =\ell^3(2\pi)^{3/2}\exp\left[-\frac{q^2\ell^2}{2}\right]
\end{equation}
which finally gives
\begin{equation}
   \frac{1}{\mell_s(\omega)}=\frac{1}{\ell_s(\omega)}=k_0^2\ell|\Delta\epsilon(\omega)|^2
      \frac{\sqrt{2\pi}}{4}\left[1-\exp(-2k_0^2\ell^2)\right].
\end{equation}

The same analysis has now to be applied to the intensity vertex $\Gamma$ and its counterpart $\mGamma$ when scatterer
displacements are present. The intensity vertex is an important quantity entering the Bethe-Salpeter equation that
describes the evolution of the field-field correlation function. It contains all possible scattering sequences for the
field and its complex conjugate counterpart that cannot be statistically factorized. Still in a dilute medium, it can be
limited to the first order which writes
\begin{equation}\label{gamma}
   \Gamma(\bm{r},\bm{r}',\bm{\uprho},\bm{\uprho}',\omega)
      =\bra V(\bm{r},\omega) V^*(\bm{\uprho},\omega)\ket_c \delta(\bm{r}'-\bm{r}')\delta(\bm{\uprho}-\bm{\uprho'}).
\end{equation}
For the case of scatterer displacements, we have
\begin{equation}\label{new_gamma}
   \mGamma(\bm{r},\bm{r}',\bm{\uprho},\bm{\uprho}',\omega)
      =\bra V(\bm{r},\omega) \mV^*(\bm{\uprho},\omega)\ket_c \delta(\bm{r}'-\bm{r}')\delta(\bm{\uprho}-\bm{\uprho'}).
\end{equation}
It is important to note that the correlation $\nC$ involves the electric field before any displacement ($E$) and its complex
conjugate counterpart after displacement ($\mE^*$). This is the reason why only the complex conjugated potential is replaced in
Eq.~(\ref{new_gamma}) compared to Eq.~(\ref{gamma}). We also note that although the potential is real, we keep the complex conjugate notation for the
sake of understanding, \ie to show that it applies to the complex conjugate field. By Fourier transforming
Eqs.~(\ref{new_gamma}) and (\ref{gamma}) and using Eqs.~(\ref{new_potential_fourier}), we obtain
\begin{equation}
   \mGamma(\bm{k},\bm{k}',\bm{\upkappa},\bm{\upkappa}',\omega)
      =\Gamma(\bm{k},\bm{k}',\bm{\upkappa},\bm{\upkappa}',\omega)
         \int\exp\left[-i(\bm{\upkappa}-\bm{\upkappa}')\cdot\bm{\Delta}\right]f(\bm{r},\bm{\Delta})\ud\bm{\Delta}.
\end{equation}
From the statistical homogeneity of the disorder, we get
\begin{align}
   \breve{\Gamma}(\bm{k},\bm{k}',\bm{\upkappa},\bm{\upkappa}',\omega)
      & =8\pi^3\delta(\bm{k}-\bm{k}'-\bm{\upkappa}+\bm{\upkappa}')
         \Gamma(\bm{k},\bm{k}',\bm{\upkappa},\bm{\upkappa}',\omega),
\\
   \breve{\mGamma}(\bm{k},\bm{k}',\bm{\upkappa},\bm{\upkappa}',\omega)
      & =8\pi^3\delta(\bm{k}-\bm{k}'-\bm{\upkappa}+\bm{\upkappa}')
         \mGamma(\bm{k},\bm{k}',\bm{\upkappa},\bm{\upkappa}',\omega),
\end{align}
which finally leads to
\begin{equation}
   \breve{\mGamma}(\bm{k},\bm{k}',\bm{\upkappa},\bm{\upkappa}',\omega)
      =\breve{\Gamma}(\bm{k},\bm{k}',\bm{\upkappa},\bm{\upkappa}',\omega)
         \int\exp\left[-i(\bm{\upkappa}-\bm{\upkappa}')\cdot\bm{\Delta}\right]f(\bm{r},\bm{\Delta})\ud\bm{\Delta}
\end{equation}
where $\breve{\Gamma}$ and $\breve{\mGamma}$ are the reduced intensity vertices. The standard phase function is given by
$\breve{\Gamma}$ through the relation
\begin{equation}
   \frac{1}{\ell_s(\omega)}p(\bm{u},\bm{u}',\omega)
      =\frac{1}{16\pi^2}\breve{\Gamma}(k_0\bm{u},k_0\bm{u}',k_0\bm{u},k_0\bm{u}',\omega).
\end{equation}
By definition, the phase function $p$ is normalized such that
\begin{equation}
   \int p(\bm{u},\bm{u}',\omega)\ud\Omega'=1.
\end{equation}
These results allow to define a generalized phase function $\mmp$ for the case where there are scatterer displacements
given by
\begin{equation}
   \frac{1}{\mell_s(\omega)}\mmp(\bm{r},\bm{u},\bm{u}',\omega)
      =\frac{1}{\ell_s(\omega)}p(\bm{u},\bm{u}',\omega)g(\bm{r},\bm{u},\bm{u}',\omega)
      =\frac{1}{16\pi^2}\breve{\mGamma}(k_0\bm{u},k_0\bm{u}',k_0\bm{u},k_0\bm{u}',\omega)
\end{equation}
where
\begin{equation}
   g(\bm{r},\bm{u},\bm{u}',\omega)=
      \int\exp\left[-i\bm{q}\cdot\bm{\Delta}\right]f(\bm{r},\bm{\Delta})\ud\bm{\Delta}
\end{equation}
and $\bm{q}=k_0(\bm{u}-\bm{u}')$ is the scattering vector. Plugging the expression of the correlation function
$C_{\epsilon}$ into Eq.~(\ref{gamma}) leads to
\begin{equation}
   p(\bm{u},\bm{u}',\omega)=\frac{k_0^2\ell^2\exp[-q^2\ell^2/2]}{2\pi\left[1-\exp(-2k_0^2\ell^2)\right]}.
\end{equation}
This concludes the computation of the building blocks required to describe light propagation in a diluted dynamic scattering medium.

\subsection{Linear regime}

We now consider the case of light transport in the linear regime. In a dilute medium such that $k_0\ell_s\gg 1$, we can show that
the field and its complex conjugate follow the same scattering sequences (after statistical average) which can be represented by the following diagram:
\begin{equation}\label{ladder}
   \begin{ddiag}{32}
      \rput(0,3){$E(\bm{r},\omega)$}
      \rput(0,-3){$\mE^*(\bm{r},\omega)$}
      \ggmoy{5}{10}{-3}
      \ggmoy{5}{10}{3}
      \ggmoy{10}{16}{-3}
      \ggmoy{10}{16}{3}
      \ggmoy{16}{22}{-3}
      \ggmoy{16}{22}{3}
      \eemoy{22}{27}{-3}
      \eemoy{22}{27}{3}
      \ccorreldeuxc{10}{3}{10}{-3}
      \ccorreldeuxc{16}{3}{16}{-3}
      \ccorreldeuxc{22}{3}{22}{-3}
      \pparticule{10}{-3}
      \pparticule{10}{3}
      \pparticule{16}{-3}
      \pparticule{16}{3}
      \pparticule{22}{-3}
      \pparticule{22}{3}
      \rput(30,3){$E_0$}
      \rput(30,-3){$E_0^*$}
   \end{ddiag}.
\end{equation}
In this representation, the top line represents a path for the electric field $E$ and the bottom line is for a path of
its complex conjugate $\mE^*$ in presence of scatterer displacements. Solid and dashed thick lines correspond to average
Green functions (describing propagation between consecutive scattering events) and average fields respectively.
Circles denote scattering events and vertical dashed lines represent statistical correlations between scattering events
through Eq.~(\ref{chi_correlation}).

This specific diagram is called the ladder and is the leading contribution to the expression of the field-field
correlation function $\nC$. Indeed the fact that it corresponds to the same path for the field and its complex conjugate
implies that there is always constructive interference between both. From this diagram, we deduce that the correlation
function $\nC$ is described by a Radiative Transfer Equation (RTE) which writes~\cite{RYTOV-1989}
\begin{equation}\label{rte}
   \left[\bm{u}\cdot\bm{\nabla}_{\bm{r}}+\frac{1}{\ell_s(\omega)}\right]\mI(\bm{r},\bm{u},\omega)
\\
      =\frac{1}{\ell_s(\omega)}\int p(\bm{u},\bm{u}',\omega)g(\bm{r},\bm{u},\bm{u}',\omega)\mI(\bm{r},\bm{u}',\omega)\ud\bm{u}'
\end{equation}
where the specific intensity $\mI$ is defined by the field-field correlation
\begin{equation}
   \delta(k-k_0)\mI(\bm{r},\bm{u},\omega)=\int \bra E\left(\bm{r}+\frac{\bm{s}}{2},\omega\right)
                                                \mE^*\left(\bm{r}-\frac{\bm{s}}{2},\omega\right)\ket
                                                e^{-ik\bm{u}\cdot\bm{s}}\ud\bm{s}.
\end{equation}

\subsection{Second harmonic regime}

We now move on to the second harmonic regime.
As stated in the main text, we apply a perturbative approach to compute the second harmonic correlation. This means that
we have first the propagation of the field at the frequency $\omega$, then second harmonic generation and finally
propagation of the field at the frequency $2\omega$. That being said, the most difficult task now is to determine the
typical pairs of paths for the field and it complex conjugate that have the leading contribution to the correlation.
This is equivalent to determine the diagrams that lead to constructive interferences. We follow the same idea than for
the linear regime thus assuming that we have essentially ladder diagrams for the beams at $\omega$ and $2\omega$.
However these ladders have to be connected by a kernel corresponding to the second harmonic generation process. This
leads to the diagram
\begin{equation}\label{non_linear_ladder}
   \begin{ddddiag}{60}
      \rput(0,3){$E(\bm{r},2\omega)$}
      \rput(0,-3){$\mE^*(\bm{r},2\omega)$}
      \ggmoy{5}{11}{-3}
      \ggmoy{11}{17}{-3}
      \ggmoy{17}{23}{-3}
      \ggmoy{23}{29}{-3}
      \ggmoy{29}{35}{-3}
      \ggmoy{35}{41}{-3}
      \ggmoy{41}{47}{-3}
      \eemoy{47}{53}{-3}
      \ggmoy{5}{11}{3}
      \ggmoy{11}{17}{3}
      \ggmoy{17}{29}{3}
      \ggmoy{29}{35}{3}
      \ggmoy{35}{41}{3}
      \ggmoy{41}{47}{3}
      \eemoy{47}{53}{3}
      \ccorreldeuxc{11}{-3}{11}{3}
      \ccorreldeuxc{17}{-3}{17}{3}
      \ccorreldeuxc{23}{-3}{23}{3}
      \ccorreldeuxc{29}{-3}{29}{3}
      \ccorreldeuxc{35}{-3}{35}{3}
      \ccorreldeuxc{41}{-3}{41}{3}
      \ccorreldeuxc{47}{-3}{47}{3}
      \pparticule{11}{-3}
      \pparticule{17}{-3}
      \pparticule{23}{-3}
      \pparticule{35}{-3}
      \pparticule{41}{-3}
      \pparticule{47}{-3}
      \pparticule{11}{3}
      \pparticule{17}{3}
      \pparticule{23}{3}
      \pparticule{35}{3}
      \pparticule{41}{3}
      \pparticule{47}{3}
      \gggmoy{29}{37}{3}{9}
      \ggmoy{37}{43}{9}
      \ggmoy{43}{49}{9}
      \eemoy{49}{55}{9}
      \ccorreldeuxc{37}{-9}{37}{-4}
      \ccorreldeuxc{37}{-2}{37}{2}
      \ccorreldeuxc{37}{4}{37}{9}
      \ccorreldeuxc{43}{-9}{43}{-4}
      \ccorreldeuxc{43}{-2}{43}{2}
      \ccorreldeuxc{43}{4}{43}{9}
      \ccorreldeuxc{49}{-9}{49}{-4}
      \ccorreldeuxc{49}{-2}{49}{2}
      \ccorreldeuxc{49}{4}{49}{9}
      \pparticule{37}{9}
      \pparticule{43}{9}
      \pparticule{49}{9}
      \gggmoy{29}{37}{-3}{-9}
      \ggmoy{37}{43}{-9}
      \ggmoy{43}{49}{-9}
      \eemoy{49}{55}{-9}
      \pparticule{37}{-9}
      \pparticule{43}{-9}
      \pparticule{49}{-9}
      \nnonlineaire{29}{-3}
      \nnonlineaire{29}{3}
      \rput(58,9){$E_0$}
      \rput(58,3){$E_0$}
      \rput(58,-3){$E_0^*$}
      \rput(58,-9){$E_0^*$}
   \end{ddddiag}
\end{equation}
where the squares denote the second harmonic processes.  We may consider that the non-linear processes occur at two
different positions for the electric field $E$ and its complex conjugate counterpart $\mE^*$. However, this would lead to a
propagation of the correlations $\bra E(\bm{r},\omega)\mE^*(\bm{r},2\omega)\ket$ or $\bra
E(\bm{r},2\omega)\mE^*(\bm{r},\omega)\ket$ which are supposed to vanish since they involve fields at two different
frequencies. This is the reason why we have a disorder correlation function $C_{\epsilon}$ between the second harmonic processses (dashed line
between the squares). Strictly speaking, we should also take into account the degeneracy of the diagram (factor $4$).
Indeed, it corresponds to all possible permutations of the incident fields. However, this will not play any role in the
following and this factor will be taken into account in the constant $\alpha$.

The kernel dressed with the two ladders at frequency $\omega$ can be considered as a source term for the second harmonic ladder. It is given by
\begin{multline}
   S(\bm{r},\bm{\uprho},2\omega)
      =\int\bra G(\bm{r}-\bm{r}',2\omega)\ket\bra G^*(\bm{\uprho}-\bm{\uprho}',2\omega)\ket
         \mGammaSHG(\bm{r}',\bm{r}'',\bm{r}''',\bm{\uprho}',\bm{\uprho}'',\bm{\uprho}''',2\omega)
\\\times
         \bra E(\bm{r}'',\omega)\mE^*(\bm{\uprho}'',2\omega)\ket
         \bra E(\bm{r}''',\omega)\mE^*(\bm{\uprho}''',2\omega)\ket
         \ud\bm{r}'\ud\bm{r}''\ud\bm{r}'''\ud\bm{\uprho}'\ud\bm{\uprho}''\ud\bm{\uprho}'''
\end{multline}
where $\mGammaSHG$ is the SHG vertex given by
\begin{equation}\label{new_gamma_shg}
   \mGammaSHG(\bm{r},\bm{r}',\bm{r}'',\bm{\uprho},\bm{\uprho}',\bm{\uprho}'',\omega)
      =\bra\chi(\bm{r})\mchi^*(\bm{\uprho})\ket_c\delta(\bm{r}-\bm{r}')\delta(\bm{r}-\bm{r}'')
                                   \delta(\bm{\uprho}-\bm{\uprho}')\delta(\bm{\uprho}-\bm{\uprho}'').
\end{equation}
Similarly to the case of $\mGamma$, we keep the complex conjugate notation for the second order susceptibility although
it is a real quantity in order remind that it corresponds to the complex conjugate field. Without any scatterer
displacement, we have
\begin{equation}\label{gamma_shg}
   \GammaSHG(\bm{r},\bm{r}',\bm{r}'',\bm{\uprho},\bm{\uprho}',\bm{\uprho}'',\omega)
      =\bra\chi(\bm{r})\chi^*(\bm{\uprho})\ket_c\delta(\bm{r}-\bm{r}')\delta(\bm{r}-\bm{r}'')
                                   \delta(\bm{\uprho}-\bm{\uprho}')\delta(\bm{\uprho}-\bm{\uprho}'').
\end{equation}
By Fourier transforming Eqs.~(\ref{new_gamma_shg}) and (\ref{gamma_shg}) and making use of Eq.~(\ref{new_chi_fourier}), we obtain
\begin{equation}
   \mGammaSHG(\bm{k},\bm{k}',\bm{k}'',\bm{\upkappa},\bm{\upkappa}',\bm{\upkappa}'',\omega)
      =\GammaSHG(\bm{k},\bm{k}',\bm{k}'',\bm{\upkappa},\bm{\upkappa}',\bm{\upkappa}'',\omega)
         \int\exp\left[-i(\bm{\upkappa}-\bm{\upkappa}'-\bm{\upkappa}'')\cdot\bm{\Delta}\right]f(\bm{r},\bm{\Delta})\ud\bm{\Delta}.
\end{equation}
From the statistical homogeneity of the disorder, we get
\begin{align}
   \tGammaSHG(\bm{k},\bm{k}',\bm{k}'',\bm{\upkappa},\bm{\upkappa}',\bm{\upkappa}'',\omega)
      & =8\pi^3\delta(\bm{k}-\bm{k}'-\bm{k}''-\bm{\upkappa}+\bm{\upkappa}'+\bm{\upkappa}'')
         \GammaSHG(\bm{k},\bm{k}',\bm{k}'',\bm{\upkappa},\bm{\upkappa}',\bm{\upkappa}'',\omega),
\\
   \tmGammaSHG(\bm{k},\bm{k}',\bm{k}'',\bm{\upkappa},\bm{\upkappa}',\bm{\upkappa}'',\omega)
      & =8\pi^3\delta(\bm{k}-\bm{k}'-\bm{k}''-\bm{\upkappa}+\bm{\upkappa}'+\bm{\upkappa}'')
         \mGammaSHG(\bm{k},\bm{k}',\bm{k}'',\bm{\upkappa},\bm{\upkappa}',\bm{\upkappa}'',\omega),
\end{align}
which finally leads to
\begin{equation}
   \tmGammaSHG(\bm{k},\bm{k}',\bm{k}'',\bm{\upkappa},\bm{\upkappa}',\bm{\upkappa}'',\omega)
      =\tGammaSHG(\bm{k},\bm{k}',\bm{k}'',\bm{\upkappa},\bm{\upkappa}',\bm{\upkappa}'',\omega)
         \int\exp\left[-i(\bm{\upkappa}-\bm{\upkappa}'-\bm{\upkappa}'')\cdot\bm{\Delta}\right]f(\bm{r},\bm{\Delta})\ud\bm{\Delta}
\end{equation}
where $\tGammaSHG$ and $\tmGammaSHG$ are the reduced SHG vertices. From this, we can define a SHG phase function given
by
\begin{equation}
   \alpha\pSHG(\bm{u},\bm{u}',\bm{u}'',\omega)=\frac{1}{125\pi^5}
      \tGammaSHG(2k_0\bm{u},k_0\bm{u}',k_0\bm{u}'',2k_0\bm{u},k_0\bm{u}',k_0\bm{u}'',\omega).
\end{equation}
$\alpha$ is a coefficient that takes into account all constants involved in the second harmonic process and is such that
the second harmonic phase function is normalized, \ie
\begin{equation}
   \int \pSHG(\bm{u},\bm{u}',\bm{u}'',\omega)\ud\Omega'\ud\Omega''=1.
\end{equation}
These results allow to define a generalized SHG phase function given by
\begin{equation}
   \alpha\mpSHG(\bm{u},\bm{u}',\bm{u}'',\omega)=\alpha\pSHG(\bm{u},\bm{u}',\bm{u}'',\omega)\gSHG(\bm{r},\bm{u},\bm{u}',\bm{u}'',\omega)
      =\frac{1}{125\pi^5}\tmGammaSHG(2k_0\bm{u},k_0\bm{u}',k_0\bm{u}'',2k_0\bm{u},k_0\bm{u}',k_0\bm{u}'',\omega)
\end{equation}
where
\begin{equation}
   \gSHG(\bm{r},\bm{u},\bm{u}',\bm{u}'',\omega)=
      \int\exp\left[-i\bm{q}\cdot\bm{\Delta}\right]f(\bm{r},\bm{\Delta})\ud\bm{\Delta}
\end{equation}
with $\bm{q}=k_0(2\bm{u}-\bm{u}'-\bm{u}'')$ the SHG scattering vector.
Plugging the expression of the correlation function $C_{\chi}$ into Eq.~(\ref{gamma_shg}) leads to
\begin{equation}
   \pSHG(\bm{u},\bm{u}',\bm{u}'',\omega)\propto\exp\left[-\frac{q^2\ell^2}{2}\right].
\end{equation}
We finally get the formulation of the RTE for the second harmonic
specific intensity linked to the second harmonic correlation function which reads
\begin{multline}\label{non_linear_rte}
   \left[\bm{u}\cdot\bm{\nabla}_{\bm{r}}+\frac{1}{\ell_s(2\omega)}\right]\mI(\bm{r},\bm{u},2\omega)
      =\frac{1}{\ell_s(2\omega)}\int p(\bm{u}\cdot\bm{u}',2\omega)g(\bm{u}\cdot\bm{u}',2\omega)\mI(\bm{r},\bm{u}',2\omega)\ud\bm{u}'
\\
      +\alpha\iint \pSHG(\bm{u},\bm{u}',\bm{u}'',\omega)\gSHG(\bm{u},\bm{u}',\bm{u}'',\omega)
         \mI(\bm{r},\bm{u}',\omega)\mI(\bm{r},\bm{u}'',\omega)\ud\bm{u}'\ud\bm{u}''.
\end{multline}
The second harmonic specific intensity if still given by the field-field correlation
\begin{equation}
   \delta(k-k_0)\mI(\bm{r},\bm{u},2\omega)=\int \bra E\left(\bm{r}+\frac{\bm{s}}{2},2\omega\right)
                                                \mE^*\left(\bm{r}-\frac{\bm{s}}{2},2\omega\right)\ket
                                                e^{-ik\bm{u}\cdot\bm{s}}\ud\bm{s}.
\end{equation}
This concludes the derivation of the RTE for the linear and second harmonic beams.

\section{Validity check from ab initio simulations}

The model developed in the previous section has been obtained under several approximations. The most important one
concerns the diagrams that have to be taken into account in order to estimate the second harmonic specific intensity. In
a dilute medium, we have considered that a ladder-type diagram is the leading contribution to the second
harmonic speckle correlation. In order to check the
validity of this approximation, we have run \emph{ab initio} simulations of Maxwell equations using a coupled-dipole
formalism and compared the results to the RTE model solved using a Monte Carlo scheme.

It is important to note that we consider here a simplified model that does not reflect the conditions of the experiment
but it will help to check the validity of the transport model. In particular, the coupled-dipole formalism used here
implies that the disorder model is limited to point scatterers randomly located inside the medium. Regarding the
scatterer displacement model, we simply consider that a scatterer can move in an arbitrary direction over a distance $d$
that is fixed.

Moreover the numerical resolution of Maxwell equations requires significant computing resources. This is the reason why we
restrict to 2D systems in TE polarization (electric field along the direction of invariance by translation). This means
that a scalar model can be used and no polarization effects have to be taken into account.

\subsection{Coupled-dipole model}

\subsubsection{Linear regime}

In the linear regime, the coupled-dipole equations are given by
\begin{align}\label{coupled_dipoles}
   E_i(\omega) & =E_0(\bm{r}_i,\omega)+k_0^2\alpha(\omega)\sum_{\substack{j=1\\j\ne i}}^N
         G_0(\bm{r}_i-\bm{r}_j,\omega)E_j(\omega),
\\\label{coupled_dipoles_field}
   E(\bm{r},\omega) & =E_0(\bm{r},\omega)+k_0^2\alpha(\omega)\sum_{j=1}^N
         G_0(\bm{r}-\bm{r}_j,\omega)E_j(\omega).
\end{align}
$E_i$ represents the field illuminating the scatterer $i$ lying at position $\bm{r}_i$. It is also called the exciting
field. It is given by two contributions: the incident field $E_0$ and the field scattered by all other scatterers.
$G_0(\bm{r}-\bm{r}_0,\omega)$ is the Green function in vacuum. It links the field created at position $\bm{r}$ by a
source dipole $p(\omega)$ lying at position $\bm{r}_0$ through the relation
\begin{equation}
   E(\bm{r},\omega) = \mu_0\omega^2 G_0(\bm{r}-\bm{r}_0,\omega)p(\omega).
\end{equation}
For 2D TE waves, it is given by
\begin{equation}
   G_0(\bm{R},\omega)=\frac{i}{4}\operatorname{H}_0^{(1)}(k_0|\bm{R}|)
\end{equation}
where $\operatorname{H}_0^{(1)}$ is the Hankel function of first kind and zero order. $\alpha(\omega)$ is the polarizability of
the scatterer. It describes the optical response of the particle. In the non-absorbing case, energy conservation
implies
\begin{equation}
   k_0\im\alpha(\omega) = \frac{k_0^3}{4}|\alpha(\omega)|^2.
\end{equation}
Once the exciting fields have been computed for all dipoles by solving the set of Eqs.~(\ref{coupled_dipoles}), the
field at any position, \ie $E(\bm{r},\omega)$, can be computed using Eq.~(\ref{coupled_dipoles_field}).

In order to take into account scatterer displacements, we just have to redo the same computation after moving the scatterers
by a distance $d$ in arbitrary directions.
Then we have access to $\mE(\bm{r},\omega)$. An average over several disorder configurations allows to estimate the
correlation function
\begin{equation}
   \nC_{\text{CD}}(\bm{r},\omega)=\frac{\bra E(\bm{r},\omega)\mE^*(\bm{r},\omega)\ket}{\bra E(\bm{r},\omega)E^*(\bm{r},\omega)\ket}
\end{equation}
where the subscript CD means ``Coupled Dipoles''.

\subsubsection{Non-linear regime}

The second harmonic regime is still considered through the standard perturbative approach. This means that the
coupled-dipole equations write
\begin{align}\label{non_linear_coupled_dipoles}
   E_i(2\omega) & =\frac{\beta}{\alpha(2\omega)}E_i(\omega)^2+k_0^2\alpha(2\omega)\sum_{\substack{j=1\\j\ne i}}^N
         G_0(\bm{r}_i-\bm{r}_j,2\omega)E_j(2\omega),
\\\label{non_linear_coupled_dipoles_field}
   E(\bm{r},2\omega) & =k_0^2\alpha(2\omega)\sum_{j=1}^N
         G_0(\bm{r}-\bm{r}_j,2\omega)E_j(2\omega)
\end{align}
where $\beta$ can be seen as a second harmonic polarizability. This set of equations is solved using the results of the
linear case and following exactly the same steps. This leads to an estimate of the correlation function
\begin{equation}
   \nC_{\text{CD}}(\bm{r},2\omega)=\frac{\bra E(\bm{r},2\omega)\mE^*(\bm{r},2\omega)\ket}{\bra E(\bm{r},2\omega)E^*(\bm{r},2\omega)\ket}.
\end{equation}

\subsection{Monte Carlo scheme}

In order to solve the set of transport equations, we have performed Monte Carlo simulations. For a cloud of uncorrelated
point dipoles lying in a 2D dilute medium and considering TE waves, the parameters are given by
\begin{align}
   \ell_s(\omega) & = \frac{1}{\rho\sigma_s(\omega)}
      \quad\text{where}\quad \sigma_s(\omega)=\frac{k_0^3}{4}|\alpha(\omega)|^2,
\\
   p(\bm{u},\bm{u}',\omega) & = \frac{1}{2\pi},
   \quad
   \pSHG(\bm{u},\bm{u}',\bm{u}'',\omega) = \frac{1}{4\pi^2}
\\
   \text{and}\quad f(\bm{r},\bm{\Delta}) & = \frac{\delta(\Delta-d)}{2\pi\Delta},
\end{align}
$\rho$ being the density of scatterers. Two different Monte Carlo simulations are done. The first one solves
Eq.~(\ref{rte}). This is a fully standard Monte Carlo scheme except that we have to multiply the energy quanta of the
random walk packets by $g(\bm{r},\bm{u},\bm{u}',\omega)$ at each scattering event in order to take into account the
decorrelation process. At the end, we have access to a map of the specific intensity $\mI(\bm{r},\bm{u},\omega)$ which
is used to compute the correlation function
\begin{equation}
   \nC_{\text{MC}}(\bm{r},\omega)=\frac{\int \mI\left(\bm{r},\bm{u},\omega\right)\ud\bm{u}}
      {\int I\left(\bm{r},\bm{u},\omega\right)\ud\bm{u}}
\end{equation}
where the subscript MC means ``Monte Carlo''.  The second Monte Carlo simulation is used to solve
Eq.~(\ref{non_linear_rte}) and is similar to the first one except that the source term is given by the SHG process [last
term of Eq.~(\ref{non_linear_rte})]. At the end, we have access to the specific intensity $\mI(\bm{r},\bm{u},2\omega)$
which gives the correlation function
\begin{equation}
   \nC_{\text{MC}}(\bm{r},2\omega)=\frac{\int \mI\left(\bm{r},\bm{u},2\omega\right)\ud\bm{u}}
      {\int I\left(\bm{r},\bm{u},2\omega\right)\ud\bm{u}}.
\end{equation}

\subsection{Numerical result comparison}

\begin{figure}[!htb]
   \centering
   \psfrag{A}[c]{(a)}
   \psfrag{B}[c]{(b)}
   \psfrag{a}[c]{$a$}
   \psfrag{z}[c]{$z$}
   \psfrag{L}[c]{$L$}
   \psfrag{d}[c]{$r$}
   \psfrag{w}[c]{$w$}
   \psfrag{D}[c]{$D$}
   \includegraphics[width=0.8\linewidth]{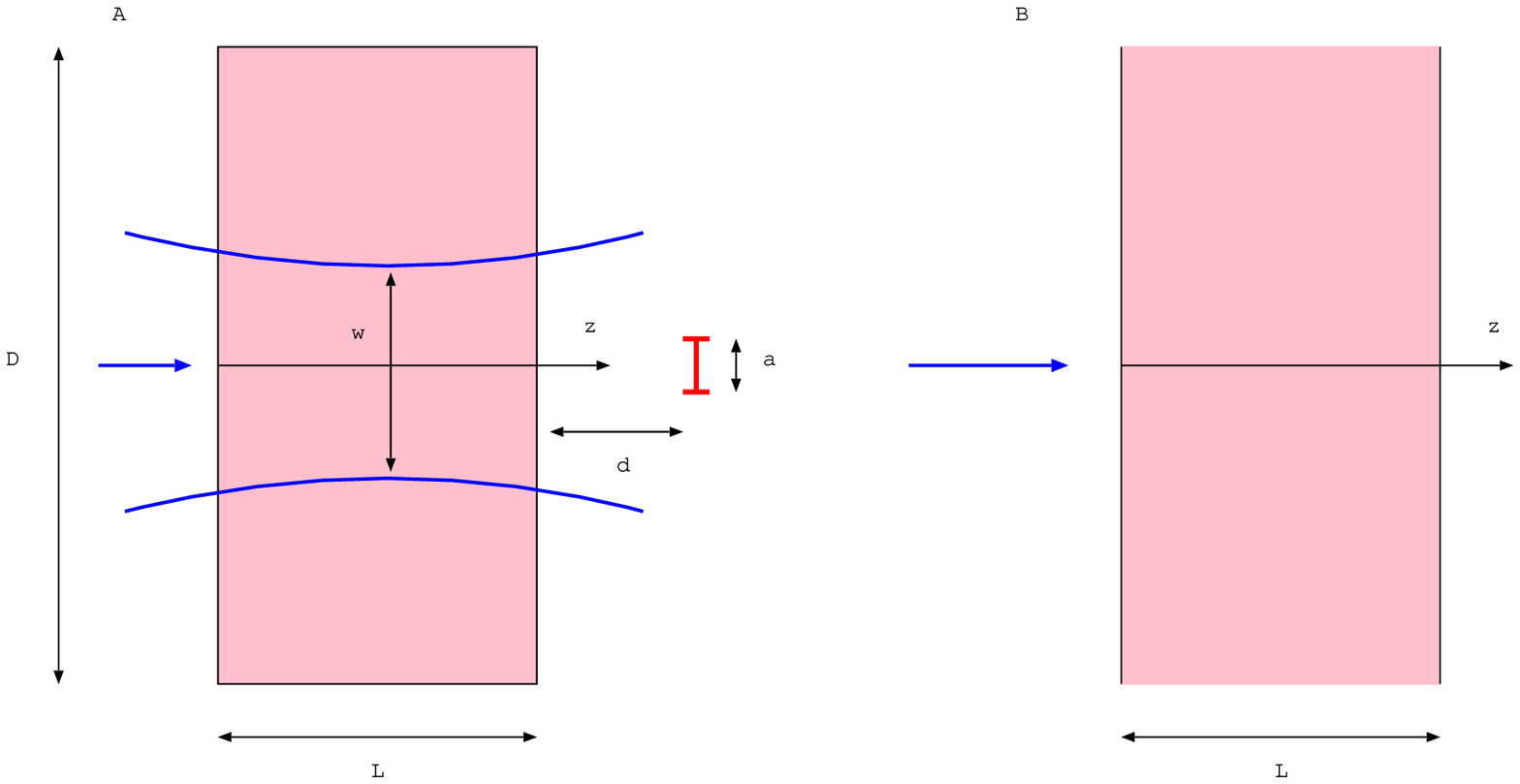}
   \caption{2D slab geometry considered to validate the transport approach. (a) Scattering rectangle used in the
   coupled-dipole simulation. We have $D\gg\{L,w\}$ and $w\gg\{L,\lambda\}$ to mimic a slab geometry shine with a
   plane-wave at normal incidence. $a$ is the transverse size of the detection zone where we compute the correlation
   functions $\nC$. We have $w\gg a\gg\lambda$ to capture several speckle grains in order to improve the statistical convergence. We
   have also $r>\lambda$ to avoid near-field interaction close to the interface. (b) Infinite scattering slab used in
   the Monte Carlo simulation. Since the \emph{ab initio} computation is done using point dipoles, the standard phase
   function $p$ as well as the SHG phase function $\pSHG$ are chosen to be isotropic ($g=0$).}
   \label{slab_num}
\end{figure}

To make comparisons between the couple-dipole formalism and the Monte carlo simulations, we consider a slab geometry of
thickness $L$ illuminated by a plane-wave at normal incidence as represented in Fig.~\ref{slab_num}\,(b). However this
very simple model have to be slightly adapted in the case of the coupled-dipole simulation which is detailed in
Fig.~\ref{slab_num}\,(a). Indeed, the finite number of scatterers imposes a finite transverse size $D$. Moreover, in
order to avoid diffraction effects at the transverse boundaries, we choose to illuminate the medium by a gaussian beam of waist $w$. We choose
$D\gg\{L,w\}$ and $w\gg\{L,\lambda\}$. The speckle correlation functions are computed just behind the slab, at a distance $r$
from the interface and average along a transverse distance $a$ in the case of the coupled dipole simulation. In the case
of the Monte Carlo simulation, we consider the full transmitted specific intensity.

\begin{figure}[!htb]
   \centering
   \psfrag{kd}[c]{\large $k_0d$}
   \psfrag{C}[c]{\large $\nC$}
   \psfrag{AAAAA}[Bl]{$\nC_{\text{CD}}(\omega)$}
   \psfrag{BBBBB}[Bl]{$\nC_{\text{CD}}(2\omega)$}
   \psfrag{CCCCC}[Bl]{$\nC_{\text{MC}}(\omega)$}
   \psfrag{DDDDD}[Bl]{$\nC_{\text{MC}}(2\omega)$}
   \includegraphics[width=0.60\linewidth]{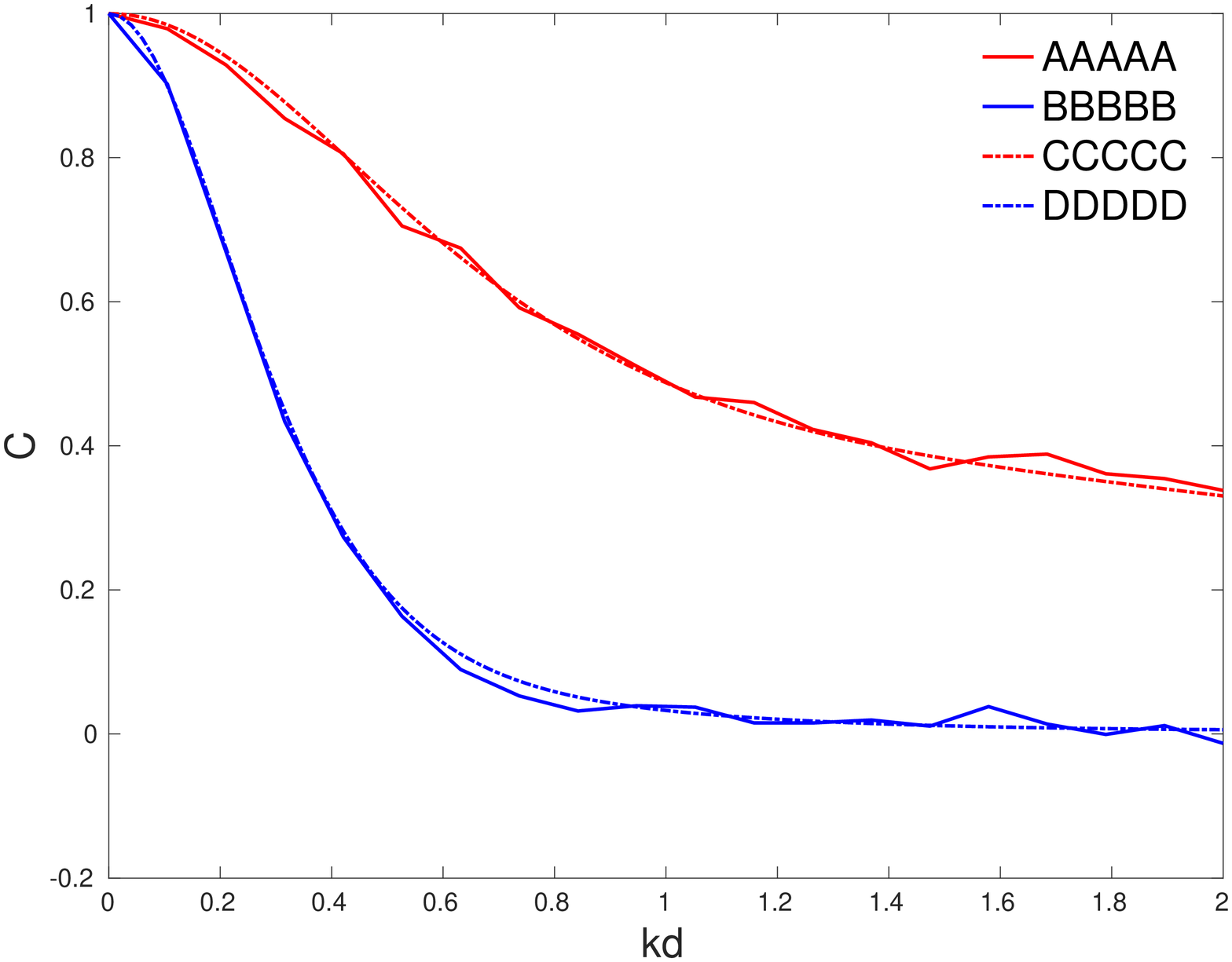}
   \caption{Comparison of the correlation functions obtained using the coupled dipole simulations ($\nC_{\text{CD}}$)
   and the Monte Carlo schemes ($\nC_{\text{MC}}$) for the linear ($\omega$) and second harmonic ($2\omega$) beams as a
   function of the normalized displacement distance $k_0d$. The parameters are $k_0L=100$, $k_0D=8000$,
   $k_0/\sqrt{\rho}=6$, $k_0w=3800$, $k_0r=5$ and $k_0a=8$. The polarizability is chosen such that
   $k_0\ell_s(\omega)=50$ and $k_0\ell_s(2\omega)=50$ which leads to $b(\omega)=b(2\omega)=2$. $N_c=256$ configurations
   are used for the coupled dipole simulations and $N_p=\num{5e6}$ random walk packets are used for the Monte Carlo
   schemes.}
   \label{comparison_cd_mc}
\end{figure}

We have tested several sets of parameters, each of them giving rise to a very good agreement between the results of both
numerical approaches. This proves that the theoretical model is accurate and that the diagram considered for the second
harmonic correlation function is the leading term. As an example, we show in Fig.~\ref{comparison_cd_mc} the results
for an optical thickness $b=L/\ell_s=2$ for both the linear and second harmonic beams.

\section{Diffusion limit}

This section is dedicated to obtaining analytical expressions of the correlation functions in the diffusion limit. We
consider a 3D slab geometry of thickness $L$ illuminated by a plane-wave at normal incidence. This implies that all
physical quantities will depend only on the depth $z$ inside the medium. The diffusion approximation requires that
$L\gg\{\ell_t(\omega),\ell_t(2\omega)\}$, $\ell_t$ being the transport mean-free path. We also assume that the
displacement direction of the scatterers is isotropic which leads to
\begin{equation}
   f(\bm{r},\bm{\Delta})=\frac{f(\Delta)}{4\pi\Delta^2}.
\end{equation}

\subsection{Linear regime}

We first consider the linear regime. Defining the absorption length $\mell_a$ and the anisotropy factor $\mg$ by
\begin{align}
   \frac{1}{\mell_a(\omega)} & =\frac{1}{\mell_e(\omega)}-\frac{1}{\mell_s(\omega)}=\frac{1}{\ell_s(\omega)}\left[1-\int
      p(\bm{u}\cdot\bm{u}',\omega)g(\bm{u}\cdot\bm{u}',\omega)\ud\bm{u}'\right],
\\
   \mg(\omega) & = \frac{1}{\mell_s(\omega)}\int \mmp(\bm{u}\cdot\bm{u}',\omega)\bm{u}\cdot\bm{u}'\ud\bm{u}'
               = \frac{\int p(\bm{u}\cdot\bm{u}',\omega)g(\bm{u}\cdot\bm{u}',\omega)\bm{u}\cdot\bm{u}'\ud\bm{u}'}
                      {\int p(\bm{u}\cdot\bm{u}',\omega)g(\bm{u}\cdot\bm{u}',\omega)\ud\bm{u}'},
\end{align}
we are back to the standard RTE with an absorption term. Thus we can apply the no less standard diffusion equation derivation in the framework
of the P1-apprximation~\cite{ISHIMARU-1997}. Since the diffusion equation is valid only for the diffuse part of the
correlation function, we split it into its ballistic ($C_b$) and diffuse ($C_d$) components which gives
\begin{equation}
   C(\bm{r},\omega)=\bra E(\bm{r},\omega)\ket\bra\mE^*(\bm{r},\omega)\ket+\bra \delta E(\bm{r},\omega)\delta \mE^*(\bm{r},\omega)\ket
      =C_b(\bm{r},\omega)+C_d(\bm{r},\omega).
\end{equation}
This splitting is similar to the one we perform usually on the average intensity. The ballistic component is given by
\begin{equation}
   C_b(z,\omega)=I_0\exp\left[-z/\ell_s(\omega)\right]
\end{equation}
and its diffuse counterpart reads
\begin{equation}
  \left[-\frac{\mell_s(\omega)\mell_t(\omega)}{3}\frac{\partial^2}{\partial
     z^2}+\frac{\mell_s(\omega)}{\mell_a(\omega)}\right]C_d(z,\omega)
   =\frac{1}{1-\mg(\omega)}C_b(z,\omega)
\end{equation}
where the transport mean-free path is given by
\begin{equation}
   \mell_t(\omega)=\frac{\mell_s(\omega)}{1-\mg(\omega)}.
\end{equation}
The boundary conditions involve the standard extrapolation length $\mz_0(\omega)=2\mell_t(\omega)/3$ and are given by
\begin{align}
   C_d(z=0,\omega)-\mz_0(\omega)\frac{\partial C_d}{\partial z}(z=0,\omega) & =
      -\frac{2\mg(\omega)}{1-\mg(\omega)}C_b(z=0,\omega),
\\
   C_d(z=L,\omega)+\mz_0(\omega)\frac{\partial C_d}{\partial z}(z=L,\omega) & =
      \frac{2\mg(\omega)}{1-\mg(\omega)}C_b(z=L,\omega).
\end{align}
The resolution of this set of equations is straightforward. Using the fact that $\ell_s\ll L$, an approximate solution
is given by
\begin{equation}
   C_d(z,\omega)=\left[-2\frac{g(\omega)}{1-g(\omega)}+3\left\{1+\frac{z_0(\omega)}{\ell_s(\omega)}\right\}\right]
      I_0\frac{\sinh[\kappa(\omega)(L-z)]}{\sinh[\kappa(\omega) L]}
      =5I_0\frac{\sinh[\kappa(\omega)(L-z)]}{\sinh[\kappa(\omega) L]}.
\end{equation}
Thus, close to the output interface in transmission, the normalized correlation is given by
\begin{equation}\label{diffuse_linear_correlation}
   \nC_d(z=L,\omega)=\frac{\kappa(\omega) L}{\sinh[\kappa(\omega) L]}
\end{equation}
where
\begin{equation}
   \kappa(\omega)=\sqrt{\frac{3}{\mell^*(\omega)\mell_a(\omega)}}.
\end{equation}
Equation~(\ref{diffuse_linear_correlation}) is the standard expression we usually obtain in the case of a large absorbing
scattering medium except that the usual absorption term describes a decorrelation process here.

\subsection{Second harmonic regime}

We now move to the second harmonic regime and we have to apply the standard diffusion equation derivation to the
non-linear RTE given by Eq.~(\ref{non_linear_rte}). Since the goal of this section is to obtain analytical results, we
have first to make an assumption on the specific intensity at frequency $\omega$ that enters the source term in the
non-linear RTE. Deeply inside the medium, we consider that its diffuse part is isotropic which gives
\begin{equation}
   \mI(\bm{r},\bm{u},\omega)=C_b(z,\omega)\delta(\bm{u}-\bm{e}_z)
      +\frac{C_d(z,\omega)}{4\pi}.
\end{equation}
The correlation at $2\omega$ is still given by
\begin{equation}
   C(\bm{r},2\omega)=\bra E(\bm{r},2\omega)\ket\bra\mE^*(\bm{r},2\omega)\ket+\bra \delta E(\bm{r},2\omega)\delta \mE^*(\bm{r},2\omega)\ket
      =C_b(\bm{r},2\omega)+C_d(\bm{r},2\omega).
\end{equation}
The ballistic component can be fully neglected. Indeed, it corresponds to a diagram where both SHG processes for the
field and its conjugate take place on statistically independent positions. This implies that phase matching cannot be
obtained and this term vanishes. Besides, it is important to note that this diagram in not taken into account in
Eq.~(\ref{non_linear_rte}).  In order to obtain a diffusion equation for the diffuse component, we first define the
first and second moments of the SHG phase function by
\begin{equation}
   \nM_0(z,\omega)=\int S(z,\bm{u},\omega)\ud\bm{u},
   \quad
   \nM_1(z,\omega)=\int S(z,\bm{u},\omega)\bm{u}\cdot\bm{e}_z\ud\bm{u}
\end{equation}
where $S$ is the source term of the RTE at $2\omega$ given by
\begin{equation}
   S(z,\bm{u},\omega)=\alpha\iint \pSHG(\bm{u},\bm{u}',\bm{u}'',\omega)\gSHG(\bm{u},\bm{u}',\bm{u}'',\omega)
      \mI(\bm{r},\bm{u}',\omega)\mI(\bm{r},\bm{u}'',\omega)\ud\bm{u}'\ud\bm{u}''.
\end{equation}
Then, we obtain the diffusion equation for the diffuse component given by
\begin{equation}
  \left[-\frac{\mell_s(2\omega)\mell_t(2\omega)}{3}\frac{\partial^2}{\partial
     z^2}+\frac{\mell_s(2\omega)}{\mell_a(2\omega)}\right]C_d(z,2\omega)
   =\mell_s(2\omega)\nM_0(z,\omega)-\mell_s(2\omega)\mell_t(2\omega)\frac{\partial}{\partial z}\nM_1(z,\omega)
\end{equation}
with the boundary conditions
\begin{align}
   C_d(z=0,2\omega)-\mz_0(2\omega)\frac{\partial C_d}{\partial z}(z=0,2\omega) & =
      -\frac{2\mell_s(2\omega)\nM_1(z=0,\omega)}{1-\mg(2\omega)},
\\
   C_d(z=L,2\omega)+\mz_0(2\omega)\frac{\partial C_d}{\partial z}(z=L,2\omega) & =
       \frac{2\mell_s(2\omega)\nM_1(z=L,\omega)}{1-\mg(2\omega)}.
\end{align}
Writing $S$ is the form
\begin{equation}
   S(z,\bm{u},\omega)=\alpha C_b(z,\omega)^2\nA(\bm{u},\omega)
      +\frac{\alpha}{4\pi} C_b(z,\omega)C_d(z,\omega)\nB(\bm{u},\omega)
      +\frac{\alpha}{16\pi^2} C_d(z,\omega)^2\nC(\bm{u},\omega)
\end{equation}
where
\begin{align}
   \nA(\bm{u},\omega) & = \pSHG(\bm{u},\bm{e}_z,\bm{e}_z,\omega)\gSHG(\bm{u},\bm{e}_z,\bm{e}_z,\omega),
\\\nonumber
   \nB(\bm{u},\omega) & = \frac{1}{4\pi}\int
      \pSHG(\bm{u},\bm{u}',\bm{e}_z,\omega)\gSHG(\bm{u},\bm{u}',\bm{e}_z,\omega)
\\ & \hphantom{=\frac{1}{4\pi}\int}
      +\pSHG(\bm{u},\bm{e}_z,\bm{u}',\omega)\gSHG(\bm{u},\bm{e}_z,\bm{u}',\omega)\ud\bm{u}'
\\ \text{and }
   \nC(\bm{u},\omega) & = \frac{1}{16\pi^2}\iint
      \pSHG(\bm{u},\bm{u}',\bm{u}'',\omega)\gSHG(\bm{u},\bm{u}',\bm{u}'',\omega)\ud\bm{u}'\ud\bm{u}'',
\end{align}
we get
\begin{align}
   \nM_0(z,\omega) & =\alpha\left[ C_b(z,\omega)^2\nA_0(\omega)
      +\frac{1}{4\pi} C_b(z,\omega)C_d(z,\omega)\nB_0(\omega)
      +\frac{1}{16\pi^2} C_d(z,\omega)^2\nC_0(\omega)\right]
\\ \text{and }
   \nM_1(z,\omega) & =\alpha\left[ C_b(z,\omega)^2\nA_1(\omega)
      +\frac{1}{4\pi} C_b(z,\omega)C_d(z,\omega)\nB_1(\omega)
      +\frac{1}{16\pi^2} C_d(z,\omega)^2\nC_1(\omega)\right]
\end{align}
with
\begin{equation}
   \nX_0(\omega)=\int \nX(\bm{u},\omega)\ud\bm{u},
   \quad
   \nX_1(\omega)=\int \nX(\bm{u},\omega)\bm{u}\cdot\bm{e}_z\ud\bm{u},
\end{equation}
where $\nX=\nA,\nB,\nC$. Since we consider that the diffusive regime is valid, we
neglect in the following the ballistic component $C_b$. Thus $\nA\sim 0$ and $\nB\sim 0$. Moreover $\nC$ is of the
form
\begin{equation}
   \nC(\bm{u},\omega) = \frac{1}{16\pi^2}\iint F(6-4\bm{u}\cdot\bm{u}'-4\bm{u}\cdot\bm{u}''+2\bm{u}'\cdot\bm{u}'',\omega)
      \ud\bm{u}'\ud\bm{u}''
\end{equation}
which implies that $\nC$ is independent of $\bm{u}$. This leads to $\nC_1=0$. Finally, only $\nC_0$ is non-sero. In
order to have its expression, we note that
\begin{equation}
   C_d(z,\omega)^2=\frac{25I_0^2}{2\sinh\left[\kappa(\omega)L\right]^2}
      \left[e^{2\kappa(\omega)(L-z)}+e^{-2\kappa(\omega)(L-z)}-2\right].
\end{equation}
Thus the diffusion equation becomes
\begin{equation}
  \left[-\frac{\mell_s(2\omega)\mell_t(2\omega)}{3}\frac{\partial^2}{\partial
     z^2}+\frac{\mell_s(2\omega)}{\mell_a(2\omega)}\right]C_d(z,2\omega)
   =\frac{25I_0\mell_s(2\omega)\alpha\nC_0(\omega)}{32\pi^2\sinh\left[\kappa(\omega)L\right]^2}
      \left[e^{2\kappa(\omega)(L-z)}+e^{-2\kappa(\omega)(L-z)}-2\right]
\end{equation}
with the boundary conditions
\begin{align}
   C_d(z=0,2\omega)-\mz_0(2\omega)\frac{\partial C_d}{\partial z}(z=0,2\omega) & =
      0,
\\
   C_d(z=L,2\omega)+\mz_0(2\omega)\frac{\partial C_d}{\partial z}(z=L,2\omega) & =
      0.
\end{align}
The resolution of this set of equations is still straightforward and we get
\begin{equation}\label{diffuse_non_linear_correlation}
   \nC_d(z=L,2\omega)
      =\frac{6\nD(\omega)}{\kappa(2\omega)L\sinh[\kappa(\omega)L]^2\sinh[\kappa(2\omega)L]}
         \frac{\kappa(2\omega)^2\left\{1-\cosh[2\kappa(\omega)L]\right\}
            -4\kappa(\omega)^2\left\{1-\cosh[\kappa(2\omega)L]\right\}}{\kappa(2\omega)^2-4\kappa(\omega)^2}
\end{equation}
where
\begin{equation}
   \nD(\omega)=\frac{\displaystyle\int
       \pSHG(\bm{u},\bm{u}',\bm{u}'',\omega)\gSHG(\bm{u},\bm{u}',\bm{u}'',\omega)\ud\bm{u}\ud\bm{u}'\ud\bm{u}''
      }{\displaystyle\int \pSHG(\bm{u},\bm{u}',\bm{u}'',\omega)\ud\bm{u}\ud\bm{u}'\ud\bm{u}''}.
\end{equation}
In this expression, we have defined
\begin{equation}
   \kappa(2\omega)=\sqrt{\frac{3}{\mell_t(2\omega)\mell_a(2\omega)}}.
\end{equation}
In the particular case where $\kappa(2\omega)=2\kappa(\omega)$, we obtain
\begin{equation}
   \nC_d(z=L,2\omega)
      =\frac{3\nD(\omega)}{\sinh[\kappa(\omega)L]}
         \left[\frac{1}{\sinh[\kappa(\omega)L]}-\frac{1}{\kappa(\omega)L\cosh[\kappa(\omega)L]}\right].
\end{equation}

\subsection{Comparison to Monte Carlo simulations}

\begin{figure}[!htb]
   \centering
   \psfrag{kd}[c]{\large $k_0d$}
   \psfrag{C}[c]{\large $\nC$}
   \psfrag{AAAAAAA}[Bl]{$\nC_{\text{MC}}(\omega)$}
   \psfrag{CCCCCCC}[Bl]{$\nC_{\text{MC}}(2\omega)$}
   \psfrag{BBBBBBB}[Bl]{$\nC_{\text{d}}(\omega)$}
   \psfrag{DDDDDDD}[Bl]{$\nC_{\text{d}}(2\omega)$}
   \includegraphics[width=0.6\linewidth]{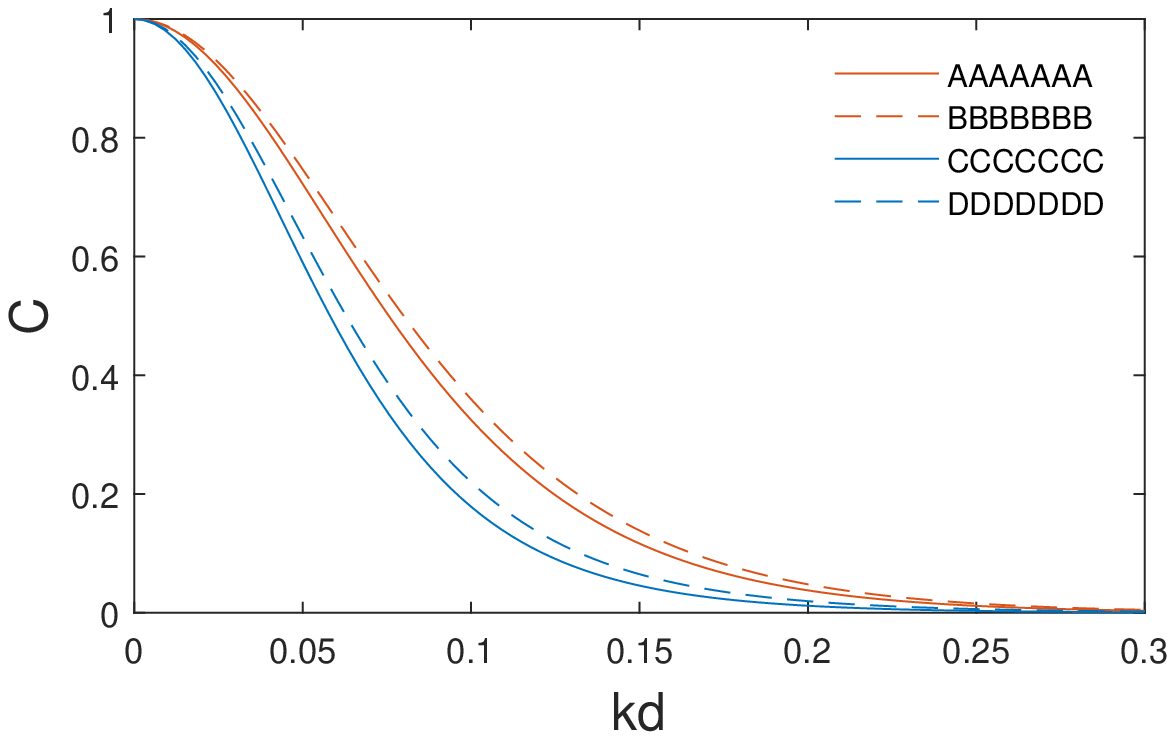}
   \caption{Comparison of the correlation functions obtained using the Monte Carlo scheme ($\nC_{\text{MC}}$)
   and diffusion approximation ($\nC_{\text{d}}$) for the linear ($\omega$) and second harmonic ($2\omega$) beams as a
   function of the normalized displacement distance $k_0d$. The parameters are $b(\omega)=b(2\omega)=40$,
   $g(\omega)=0.31$, $g(2\omega)=0.75$ and $k_0\ell=1$. $N_p=\num{28e6}$ random walk packets are used for the Monte
   Carlo schemes.}
   \label{comparison_mc_diff}
\end{figure}

In order to check the validity of Eqs.~(\ref{diffuse_linear_correlation}) and (\ref{diffuse_non_linear_correlation}), we
have performed Monte Carlo simulations in the particular case of a constant displacement amplitude $d$ such that
$k_0d\ll 1$ and $f(\Delta)=\delta(\Delta-d)$. This gives
\begin{align}
   \kappa(\omega)L & =b(\omega)k_0d,
   \quad
   \kappa(2\omega)L=2b(2\omega)k_0d
\\
   \text{and } \nD(\omega) & = \frac{1}{8\pi}\int_{\phi=0}^{2\pi}\int_{\mu=-1}^1\int_{\mu'=-1}^1
         \operatorname{sinc}\left[k_0d
            \vphantom{\sqrt{6-4\mu-4\mu'+2\mu\mu'+2\sqrt{1-\mu^2}\sqrt{1-\mu'^2}\cos\phi}}
         \sqrt{6-4\mu-4\mu'+2\mu\mu'+2\sqrt{1-\mu^2}\sqrt{1-\mu'^2}\cos\phi}\right]
            \ud\mu\ud\mu'\ud\phi.
\end{align}
An example of the results obtained is reported in Fig.~\ref{comparison_mc_diff} for $b(\omega)=b(2\omega)=40$,
$g(\omega)=0.31$ and $g(2\omega)=0.75$. The disorder correlation is given by $k_0\ell=1$. A good agreement is obtained.

\bibliography{shg_si}